\documentclass[conference]{IEEEtran}
\usepackage[utf8]{inputenc}
\usepackage[T1]{fontenc}
\usepackage{amsmath,amssymb}
\usepackage{bm}
\usepackage{booktabs}
\usepackage{graphicx}
\usepackage{multirow}
\usepackage{xcolor}
\usepackage[colorlinks=false,pdfborder={0 0 0}]{hyperref}
\hypersetup{pdftitle={How Far Back Should a Transformer Look? Repetition and Copying in Music Sequence Models}, pdfauthor={Amir Fathi}}
\usepackage{enumitem}
\usepackage{cite}
\usepackage{url}
\usepackage{placeins}
\usepackage{pgfplots}
\usetikzlibrary{patterns}
\pgfplotsset{compat=1.18}

\input{tables/numbers}

\begin{document}

\title{How Far Back Should a Transformer Look?\\Repetition and Copying in Music Sequence Models}

\author{
\IEEEauthorblockN{Amir Fathi}
\IEEEauthorblockA{Department of Computing Science\\University of Alberta\\Edmonton, AB, Canada\\fathi5@ualberta.ca}
}

\maketitle

% ============================================================
\begin{abstract}
We investigate how predictive performance depends on the maximum context available to an autoregressive model of symbolic music, and what information long-context models exploit. A small causal Transformer is trained separately at each context length $T \in \{6,18,48,96,192,336\}$ and evaluated on the same target positions, both over raw tokens and over non-overlapping binary latent codes.
On Nottingham folk tunes, the token predictor's test NLL decreases by \R{nott/tok/drop}\% (\R{nott/boot/t/6-336} bits per token) between 6 and 336 tokens, and by \R{oneill/tok/drop}\% on the O'Neill's tunes long enough for the same sweep (\R{oneill/long/test} of \R{oneill/pieces/test} test tunes). Most of the long-context gain is explained by exact copying: the improvement appears when an earlier occurrence of the target's 16-token history enters the available context; overwriting that occurrence removes the gain, whereas equally large unrelated corruption does not; and a simple copy baseline recovers \R{nott/copybase/backoff/gain}\% of the reduction. A trained 336-token model likewise loses most of this gain when its history is restricted to recent tokens at test time. Because many repeats arise from written repeat signs expanded in the score, this result pertains specifically to these rendered score representations.
By contrast, on MAESTRO performances and MusicNet scores, where exact repeats are substantially less frequent, the reduction is smaller (\R{maes/tok/drop}\% and \R{music/tok/drop}\%) and is largely attained by 96 tokens.
Finally, an audit of an earlier draft that reported saturation at 16 tokens identified split leakage, overlapping latent receptive fields, an averaging predictor, and a per-file tempo grid; we present these as methodological checks for context-length measurements.
\end{abstract}


% ============================================================
\section{Introduction}
\label{sec:intro}
% ============================================================

Autoregressive models of symbolic music commonly attend over the full available history of a piece~\cite{huang2018music,qu2024mupt,wang2025notagen}, while full self-attention incurs $O(n^2)$ cost in sequence length $n$. Restricting the model to a local context can therefore substantially reduce computation, and some systems are motivated by the premise that recent history contains most of the information relevant to next-event prediction~\cite{thickstun2024anticipatory}. This raises two empirical questions: how does predictive performance vary with the amount of available history, and which properties of the musical representation determine whether longer context is useful?

We address these questions with a controlled sweep over maximum context length. A predictor is trained and evaluated with exactly the preceding $T$ tokens, for $T$ from 6 to 336, and all context lengths are scored on the same target positions. The intended experimental factor is therefore the maximum accessible history, although the attention bands and relative-position parameters necessarily grow with $T$. Because a separate model is trained at each $T$, the sweep measures the performance attainable under a fixed training protocol at each maximum context length; by itself, it does not establish how much context any single trained model actually uses. We therefore add two interventions on trained long-context models: targeted context editing and test-time restriction of the available history. We conduct the sweep for two predictors, one over raw tokens and one over binary autoencoder codes whose latent steps each represent a disjoint span of 6 tokens. The latter reproduces the two-stage setting of our earlier draft, in which prediction operates on a latent sequence 6 times shorter than the token sequence. Both predictors use the same base causal-Transformer architecture and are trained on source-disjoint splits that keep each tune or performance entirely within one split.

Our contributions are the following.

\begin{enumerate}[nosep,leftmargin=*]
  \item \textbf{A controlled sweep of available context.} On the two folk corpora, Nottingham and O'Neill's, the token predictor's test NLL decreases by \R{nott/tok/drop}\% and \R{oneill/tok/drop}\% from 6 to 336 tokens (\R{nott/boot/t/6-336} and \R{oneill/boot/t/6-336} bits per token), with most of the improvement occurring beyond 48 tokens. On MAESTRO and MusicNet, the reductions are smaller (\R{maes/tok/drop}\% and \R{music/tok/drop}\%, corresponding to \R{maes/boot/t/6-336} and \R{music/boot/t/6-336} bits), and most of the improvement within the tested range is attained by 96 tokens (Sections~\ref{sec:sweep},~\ref{sec:oneills} and~\ref{sec:musicnet}). We do not evaluate contexts longer than 336 tokens.
  \item \textbf{Evidence that the long-context gain on the folk corpora is primarily exact copying of repeated passages.} Targets whose 16-token histories recur within the available context account for most of the NLL reduction. Performance improves on such targets only after the earlier occurrence enters the model's receptive field; selectively overwriting the earlier occurrence removes the gain; a simple copy baseline recovers most of the reduction; and a trained 336-token model loses the gain when its history is restricted at test time. The same pattern occurs on Nottingham and O'Neill's (Sections~\ref{sec:lag},~\ref{sec:copy},~\ref{sec:ablate} and~\ref{sec:oneills}). Because repeat signs are expanded into note-for-note repeated passages in both folk corpora, and the representation retains only the top line on a fixed grid, exact repetition reflects both musical structure and representation. Our claim is therefore specific to these rendered score representations.
  \item \textbf{An audit of an earlier context-saturation result.} An earlier draft reported the opposite qualitative conclusion, namely saturation at 16 tokens. We identify the methodological failures associated with that curve and formulate them as checks for future measurements of effective context length (Section~\ref{sec:correction}).
\end{enumerate}

Additional analyses show that the Nottingham token predictor acquires the copying solution abruptly and seed-dependently at 96 tokens (Section~\ref{sec:stability}); that the averaging predictor used in the earlier draft fails to exploit long context (Section~\ref{sec:v5}); and, as a secondary diagnostic, that long-context models generate passages that recur within a stream, although no generated set matches held-out music under our distance (Section~\ref{sec:gen}).

% ============================================================
\section{Related Work}
\label{sec:related}
% ============================================================

\textbf{Context in sequence models of music.} Music Transformer~\cite{huang2018music} applies relative attention over full sequences, Perceiver AR~\cite{hawthorne2022perceiverar} extends symbolic-music context to thousands of tokens, and MuPT~\cite{qu2024mupt} and NotaGen~\cite{wang2025notagen} scale language-modeling approaches to symbolic scores. MusicVAE~\cite{roberts2018hierarchical} models long-term structure with a recurrent hierarchical decoder conditioned on a global latent vector. The Anticipatory Music Transformer~\cite{thickstun2024anticipatory} is motivated by the premise that recent history is most relevant to the next event, but does not directly quantify the required history length. More broadly, WaveNet~\cite{oord2016wavenet} and temporal convolutional networks~\cite{bai2018empirical} demonstrate that finite receptive fields can be competitive with recurrent or effectively unbounded alternatives. Context utilization has been studied explicitly in language modeling: Khandelwal et al.~\cite{khandelwal2018sharp} truncate or perturb LSTM histories and report useful context extending to roughly 200 tokens, with neural caches improving the use of distant lexical repetitions; Dai et al.~\cite{dai2019transformerxl} formalize effective context length; and Sun et al.~\cite{sun2021longrange} report that additional long-range Transformer context beyond 2K tokens primarily benefits tokens that can be copied from distant history. Our experiments pose an analogous question for symbolic music, but train a separate model at each maximum context length rather than shortening the history of a single model alone.

Two recent studies are particularly relevant. Yi et al.~\cite{yi2026depth} introduce a diagnostic for functional long-range dependence and report that full-horizon models on MAESTRO exploit context beyond a local window of 1024 tokens. Ghoshal et al.~\cite{ghoshal2026fusing} compare LSTMs and Transformers on folk music and report that LSTMs capture local regularities but lose long-range structure. Our folk-corpus results are consistent with the latter observation that long context is useful, while further localizing the gain, for these rendered scores, to exact repetition of extended passages. Section~\ref{sec:discussion} relates our MAESTRO results to the former study.

\textbf{Copying.} Transformer language models can acquire copying circuits, often described as induction heads, abruptly during training, and these circuits account for a substantial fraction of long-context behavior in some settings~\cite{olsson2022induction}. Copying in our token predictors also emerges abruptly and depends on the training seed (Section~\ref{sec:stability}). Our analysis intervenes on the input context rather than interpreting attention patterns, so the connection to induction heads is an analogy rather than an identification. In symbolic music generation, Yin et al.~\cite{yin2022copies} show that a Music Transformer trained beyond its conventional stopping point can reproduce excerpts of training pieces.

\textbf{Repetition in music.} Musical repetition occurs at multiple scales, from local motifs to entire sections, and is central to perceived structure~\cite{margulis2014repeat}. Self-similarity matrices~\cite{foote1999selfsimilarity} compare pairs of time positions in a piece, with repeated sections appearing as lines parallel to the main diagonal; they are a standard basis for structural analysis. Our repeat-lag statistic can be viewed as a one-dimensional analogue: for each prediction target, it records the distance to the nearest earlier occurrence of the target's 16-token history.

\textbf{Latent representations.} VAEs~\cite{kingma2013auto} and VQ-VAE~\cite{oord2017neural} compress sequences into compact latents, and MusicVAE~\cite{roberts2018hierarchical} and Jukebox~\cite{dhariwal2020jukebox} apply these ideas to symbolic music and audio. Our autoencoder is binary, trained with a straight-through estimator~\cite{bengio2013ste}.

\textbf{Evaluation.} Yang and Lerch~\cite{yang2020evaluation} argue that metrics for music generation should be interpreted relative to values observed on real data, while Frechet Music Distance~\cite{retkowski2024fmd} measures distributional distance between generated and reference sets. Yu et al.~\cite{yu2026bach} compare autoregressive, latent-variable, and adversarial approaches for Bach-style symbolic piano generation.

% ============================================================
\section{Data}
\label{sec:data}
% ============================================================

\subsection{Corpora and splits}

We use Nottingham~\cite{boulanger2012modeling}, a corpus of approximately one thousand folk tunes, and MAESTRO v3~\cite{hawthorne2019maestro}, a corpus of recorded piano performances. For Nottingham, we use the Jukedeck MIDI release~\cite{jukedeck_nottingham}, which cleans the ABC scores of the Nottingham Music Database and renders each tune in three forms: full, melody-only, and chords-only. We retain a single rendering per tune, the melody-only MIDI file. The renderer expands written repeat signs, causing repeated sections to recur note for note. For MAESTRO, we use the official split, which keeps all performances of a composition on the same side of the split.

To assess whether the Nottingham result is corpus-specific, we add a second folk corpus: the Irish tunes from O'Neill's \emph{Music of Ireland: Eighteen Hundred and Fifty Melodies}~\cite{oneill1903}, published in 1903 and available in the public domain, using the Norbert Paap ABC transcription distributed with music21~\cite{cuthbert2010music21}. ABC folk transcriptions have previously been modeled directly with recurrent networks~\cite{sturm2016folkrnn}. We expand each tune's repeat signs with music21, matching the rendered-repeat convention of Nottingham, and exclude the \R{oneill/unusable} tunes that cannot be parsed or contain fewer than 96 steps. These dances and airs are shorter than the Nottingham tunes: \R{oneill/long/test} of \R{oneill/pieces/test} O'Neill's test tunes are long enough to provide a complete 384-token window, compared with 133 of 154 Nottingham test tunes.

Nottingham is split by tune in a 70/15/15 ratio using a fixed seed. Before splitting, tunes sharing at least half of the non-constant 16-token patterns of the smaller tune are grouped together, which captures cases in which the same tune appears under different names. The rule is applied again after tokenization (Section~\ref{sec:tok}); \R{nott/moved} tunes whose merged group would cross split boundaries are reassigned to a single split. Under this criterion, no validation or test tune is a near-duplicate of a training tune. The proportion of non-constant 48-token test blocks that also occur verbatim in training is \R{nott/leak/test/verbatim}\% for Nottingham, \R{oneill/leak/test/verbatim}\% for O'Neill's, \R{music/leak/test/verbatim}\% for MusicNet, and \R{maes/leak/test/verbatim}\% for MAESTRO (Table~\ref{tab:data}). O'Neill's is split using the same procedure, which merges \R{oneill/merged} groups under the near-duplicate rule. Its \R{oneill/ov/verbatim} training-overlapping blocks (of \R{oneill/ov/blocks}) arise from \R{oneill/ov/tunes} test tunes whose closest training tune shares \R{oneill/ov/min}\% to \R{oneill/ov/max}\% of its 16-token patterns, below the 50\% grouping threshold. Excluding these tunes changes the 6-to-336-token reduction of the O'Neill's token predictor only from \R{oneill/ov/drop/all}\% to \R{oneill/ov/drop/without}\%. Training and evaluation windows are extracted only after the split is fixed.

To provide an intermediate corpus between the folk scores and MAESTRO performances, we also use MusicNet~\cite{thickstun2017musicnet}, comprising \R{music/pieces/all} aligned MIDI scores of classical solo and chamber recordings. Files are grouped by composer and movement so that all recordings of a movement remain together, and groups sharing at least half of their 16-token patterns are merged as in Nottingham, leaving \R{music/groups} groups. A fixed-seed (42) 70/15/15 group split yields \R{music/pieces/train}, \R{music/pieces/val}, and \R{music/pieces/test} files for training, validation, and test, with no near-duplicate pair crossing a split. Tokenization uses each file's own tempo map, giving 4 steps per quarter note rather than MAESTRO's fixed 125\,ms grid. The vocabulary contains 48 tokens (Section~\ref{sec:tok}), and \R{music/clamped}\% of steps are clamped to its pitch range. We use non-overlapping 480-token windows with the final 144 tokens as targets and apply no transposition. Only token predictors are trained for MusicNet, because no MusicNet autoencoder was trained; the resulting test split contains \R{music/win/test} windows.

\begin{table*}[t]
\centering
\caption{The four corpora after tokenization. Windows are the training and evaluation windows of the context sweep (Section~\ref{sec:protocol}). The last column is the share of non-constant 48-token test blocks that occur word for word in the training split.}
\label{tab:data}
% generated by scripts/make_paper_tables.py; do not edit
\begin{tabular}{lcccccccc}
\toprule
 & \multicolumn{3}{c}{\textbf{Pieces}} & \multicolumn{3}{c}{\textbf{Tokens}} & \textbf{Windows} & \textbf{Test blocks} \\
\cmidrule(lr){2-4}\cmidrule(lr){5-7}
 & Train & Val & Test & Train & Val & Test & train / val / test & \textbf{in train} \\
\midrule
Nottingham & 729 & 150 & 154 & 382K & 90K & 84K & 2789 / 809 / 660 & 0.0\% \\
O'Neill's & 1397 & 295 & 302 & 738K & 115K & 158K & 7121 / 687 / 1505 & 0.6\% \\
MusicNet & 221 & 54 & 48 & 577K & 141K & 116K & 1091 / 267 / 218 & 0.0\% \\
MAESTRO & 962 & 137 & 177 & 4586K & 560K & 575K & 9072 / 1094 / 1113 & 0.2\% \\
\bottomrule
\end{tabular}

\end{table*}

\subsection{Tokenization}
\label{sec:tok}

Each MIDI file is quantized to 4 steps per quarter note according to its tempo map, so a 4/4 bar contains 16 grid steps. MAESTRO files contain a single tempo of 120 BPM, making the MAESTRO grid a fixed 125\,ms. At each step we retain the highest sounding pitch, following the skyline rule used in melody matching~\cite{uitdenbogerd1999melodic}. Onsets and offsets are rounded to the nearest grid step, with every note retained for at least one step. Event-based representations encode performances with note-on, note-off, and time-shift events~\cite{oore2018feeling}, or scores with note, duration, and bar-position events~\cite{huang2020remi}; consequently, a fixed number of tokens can span a variable amount of musical time. The fixed grid avoids that confound when context length is varied in tokens.

Token 0 denotes rest. Tokens 1 to 47 correspond to the 47 consecutive pitches with the greatest occupancy in the training split: starting at MIDI \R{nott/pitchlo} for Nottingham, MIDI \R{oneill/pitchlo} for O'Neill's, MIDI \R{music/pitchlo} for MusicNet, and MIDI \R{maes/pitchlo} for MAESTRO. Pitches outside the retained range are clamped to the nearest boundary, affecting no Nottingham or O'Neill's steps, \R{maes/clamped}\% of MAESTRO steps, and \R{music/clamped}\% of MusicNet steps. Because clamping can merge distinct extreme pitches and thereby introduce artificial exact matches, we report MAESTRO token predictors retrained with an 89-token vocabulary consisting of rest plus MIDI pitches 21 to 108, eliminating clamping. The latent predictor and autoencoder remain on the 48-token representation. We did not retrain MusicNet with 89 tokens; instead, we recomputed repeat frequencies on unclamped pitches. At 336 tokens, \R{music/clamp/rep16/full}\% of test targets have a 16-token repeat in view without clamping versus \R{music/clamp/rep16/clamped}\% with clamping (and \R{music/clamp/rep8/full}\% versus \R{music/clamp/rep8/clamped}\% for 8-token histories), indicating that clamping does not materially create repeats in MusicNet.

The resulting representation is monophonic: it retains only the uppermost active pitch. Consequently, the MAESTRO experiments characterize the upper voice of each piano performance rather than the full polyphonic performance. Because the representation contains no onset symbol, a sustained pitch and repeated attacks of the same pitch map to the same token sequence.

\subsection{Transposition}

Nottingham contains 729 training tunes, and the predictors overfit this set within a few dozen epochs. We therefore apply online pitch transposition to every Nottingham training window: each time a window is sampled, it is shifted by a uniformly selected interval from $-5$ to $+6$ semitones among those that keep all pitches within the vocabulary. Rest tokens remain unchanged and pitches never wrap across the vocabulary boundaries. Validation and test windows are not transposed. Key transposition has previously been used for Nottingham~\cite{walder2016beyond}. O'Neill's, with \R{oneill/long/train} training tunes long enough to produce a window, uses the same augmentation. MAESTRO provides \R{maes/win/train} training windows and MusicNet \R{music/win/train}, compared with Nottingham's \R{nott/win/train}; neither is transposed.

% ============================================================
\section{Models}
\label{sec:models}
% ============================================================

\subsection{Autoencoder with disjoint spans}
\label{sec:ae}

The autoencoder maps each 48-token block to 8 latent steps of $D{=}64$ binary variables. Latent step $j$, for $j=0,\dots,7$, is computed exclusively from tokens $6j$ through $6j{+}5$, which we verify by gradient inspection. The encoder embeds tokens in 48 dimensions, applies a stride-6, width-6 convolution to 256 channels, then two width-1 convolutions (to 256 and then $D$ channels) and two width-1 residual blocks with leaky-ReLU activations. The decoder mirrors this structure with a stride-6, width-6 transposed convolution, so latent step $j$ also reconstructs only its corresponding six-token span, producing a categorical distribution over the 48-token vocabulary at each position. The continuous encoder output is bounded, perturbed during training, and binarized as follows:
\begin{equation}
\begin{aligned}
\mathbf{z} &= \operatorname{sign}\bigl(\tanh(\mathbf{z}_{\text{cont}}) + 0.05\,\bm{\epsilon}\bigr) \in \{-1,+1\}^{D \times 8},\\
\epsilon_{ij} &\overset{\text{iid}}{\sim} \mathcal{N}(0,1),
\end{aligned}
\end{equation}
where $\mathbf{z}_{\text{cont}} \in \mathbb{R}^{D \times 8}$ denotes the encoder output, $\operatorname{sign}$ is applied elementwise, and $\epsilon_{ij}$ are independent standard normal variables. Noise is disabled at inference, and gradients are propagated through the sign operation using a straight-through estimator~\cite{bengio2013ste}. The objective is token reconstruction cross-entropy plus $\beta$ times the mean of $\tanh(\mathbf{z}_{\text{cont}})^2$ over all $D \times 8$ elements. We anneal $\beta$ linearly from $10^{-3}$ to 0 and add Gaussian noise with SD 0.1 to token embeddings during training. Optimization uses Adam with learning rate $3\times10^{-4}$ and batch size 256; the checkpoint with minimum validation loss is retained.

The autoencoder reduces sequence length by a factor of six but does not constitute bit-level compression: each latent step contains 64 bits, whereas six tokens from a 48-symbol vocabulary require only $6\log_2 48 \approx 33.5$ bits under a fixed-length code. We therefore refer to the representation as a sequence of binary latent steps rather than as a compressed bitstream.

Disjoint receptive fields are essential to the context-length analysis. In the earlier encoder, each latent step depended on approximately 22 overlapping tokens, allowing a predictor of step $j$ to access part of the very token span represented by $j$ through neighboring latent steps. Section~\ref{sec:correction} analyzes this failure mode.

The Nottingham autoencoder is trained for 200 epochs with the same transposition used for the predictors and reconstructs \R{nott/ae/acc}\% of test tokens. The O'Neill's autoencoder uses the same settings and transposition for 200 epochs with seed 42, reaching \R{oneill/ae/acc}\% test-token accuracy. The MAESTRO autoencoder is trained for 30 epochs and reaches \R{maes/ae/acc}\%. By comparison, an autoencoder using the earlier Nottingham configuration, $D{=}32$ with 128 channels, achieved only \R{maesd32/ae/acc}\% on MAESTRO. The main context sweep uses one autoencoder per corpus, so the reported predictor-seed variation does not include variation in the learned representation; Section~\ref{sec:aeseeds} quantifies that source of variation using two additional Nottingham autoencoders.

\subsection{Predictors with an exact context}
\label{sec:causal}

All predictors use a two-layer causal Transformer with pre-normalization blocks, hidden width 384, four attention heads, a feed-forward width of 768 with GELU activations, and dropout 0.2. Attention scores include learned relative-position terms, one vector per offset within each layer's attention band, so the parameter count increases slightly with context length. The representation at the most recent position is passed through a two-layer output head (384 to 192 to the output dimension). The 336-token token predictor contains \R{params/tok} parameters, and all predictors in the sweep contain between 2.47M and 2.50M parameters.

Attention is band-limited in each layer: the first layer attends to the most recent $b_1$ positions and the second to the most recent $b_2$, with $b_1+b_2-1=T$. The output at position $i$ predicts $x_{i+1}$ and can depend on exactly $x_{i-T+1},\dots,x_i$. We verify this receptive field by gradient inspection for every $T$ used in the sweep.

The first-layer band is deliberately short, at most 18 tokens or 3 latent steps, while the second layer spans the remaining history. This asymmetry is important for retrieval: copying a passage from lag $l$ requires the first layer to encode the local matching pattern and the second layer to attend directly to the corresponding distant position. In an earlier implementation with equal attention bands, a 336-token model could effectively copy only from approximately the nearest half of its nominal context; on Nottingham, performance improved strongly for repeats up to 168 tokens back but much less beyond that lag.

The latent predictor reads the $\pm1$ bits of each latent step through a linear layer and emits one logit per bit, with the bits of a step independent given the context:
\begin{equation}
p(\mathbf{z}_t \mid \mathbf{z}_{t-k:t-1}) = \prod_{d=1}^{D} \sigma\bigl(z_{t,d}\, [f_\theta(\mathbf{z}_{t-k:t-1})]_d\bigr),
\end{equation}
where $\mathbf{z}_t \in \{-1,+1\}^D$ is the latent step being predicted, $\mathbf{z}_{t-k:t-1}$ denotes the preceding $k$ steps, $f_\theta$ is the Transformer, $[\cdot]_d$ selects its $d$-th output logit, and $\sigma$ is the logistic function. A context of $k$ latent steps corresponds to $T=6k$ input tokens. The token predictor instead embeds discrete tokens and outputs a categorical distribution over 48 symbols, or 89 symbols in the unclamped MAESTRO rerun.

\subsection{The averaging predictor of our earlier draft}
\label{sec:v5model}

For comparison, we also train v5, the predictor used in the earlier draft. It consists of one attention layer with relative-position representations~\cite{shaw2018selfattention} over the $k$ context steps; the resulting positionwise representations are averaged before a feed-forward prediction head (\R{params/v5} parameters; hidden width 512). Because the model has no absolute positional encoding and uses mean pooling over context positions, it has no direct mechanism for privileging the most recent or a specific distant position.

% ============================================================
\section{Protocol}
\label{sec:protocol}
% ============================================================

\textbf{Windows and targets.} For Nottingham and O'Neill's, each evaluation window contains 8 blocks, or 384 tokens, and the final 48 tokens are prediction targets. The longest context, 336 tokens, is exactly the history available to the first target, ensuring that every tested context length is trained and evaluated on identical target positions. Windows begin every 48 tokens within a piece and are never padded; consequently, pieces shorter than 384 tokens are excluded from all context lengths rather than selectively from the longer ones. For MAESTRO and MusicNet, windows contain 480 tokens, are non-overlapping, and use the final 144 tokens as targets. The latent predictor predicts the corresponding 8 or 24 latent steps.

\textbf{Contexts.} We evaluate $T \in \{6,18,48,96,192,336\}$ input tokens, corresponding to $k \in \{1,3,8,16,32,56\}$ latent steps. The $T{=}0$ token baseline is the empirical unigram distribution of training targets; the $k{=}0$ latent baseline uses the marginal probability of each bit.

\textbf{Training.} Predictors are optimized with AdamW~\cite{loshchilov2019adamw}, learning rate $10^{-3}$, weight decay $10^{-4}$, batch size 32, gradient-norm clipping at 1.0, and label smoothing~\cite{szegedy2016inception} of 0.06 in the training objective. Training stops after 20 epochs without improvement in validation NLL, and the checkpoint with the lowest validation NLL is retained. The default cap is 150 epochs. The \R{cap/n} runs that reached this cap were still improving and were therefore retrained with a 400-epoch cap; all reported values for these configurations use the longer reruns. \R{cap/small} reruns changed by less than 0.01 bits. One 18-token rerun finished \R{cap/maxloss} bits worse despite the same nominal seed, consistent with nondeterministic GPU execution inducing a different optimization trajectory. O'Neill's uses a 400-epoch cap throughout and no run reaches it; no MusicNet or MAESTRO run reaches 150 epochs. Each configuration is trained with seeds 42, 7, and 0.

\textbf{Metric.} Test NLL is computed without label smoothing and reported as bits per latent bit for latent predictors and bits per token for token predictors. These quantities are defined on different random variables and are therefore not numerically comparable; we compare only the shapes of the context-length curves. The predictors also condition on different information within each six-token span: the latent predictor predicts an entire latent step from preceding steps only, whereas the token predictor observes earlier tokens within the same six-token interval. We use the latent sweep to test whether the qualitative context effect persists after a sixfold reduction in sequence length, not to compare model quality. Each reported NLL is the value attained under the specified training protocol at that context length; Section~\ref{sec:stability} shows that the region from 48 to 96 tokens on Nottingham depends on optimization seed and stopping rule as well as context. We report the mean and sample SD across three training seeds. For differences between context lengths, we additionally report 95\% paired-bootstrap intervals~\cite{efron1993bootstrap} over test pieces containing at least one evaluation window (\R{nott/pieces/testwin} Nottingham tunes, \R{oneill/pieces/testwin} O'Neill's tunes, \R{music/pieces/testwin} MusicNet pieces, and \R{maes/pieces/testwin} MAESTRO performances; 2000 bootstrap draws), applied to the seed-mean NLL of each target. Resampling is performed at the piece level because targets within a piece are dependent. These intervals quantify variation induced by the sampled test pieces conditional on the trained models; training-seed variation is summarized separately by the reported SD.

\textbf{Repeat lag.} Let $x_1, \dots, x_L$ be the tokens of a window, $x_t$ a target, and $h_t = (x_{t-16}, \dots, x_{t-1})$ its 16-token history. The repeat lag of the target is
\begin{equation}
\label{eq:lag}
\ell_t = \min \{\, t - j \;:\; 17 \le j < t,\ h_j = h_t \,\},
\end{equation}
the distance to the nearest earlier position $j$ whose 16-token history matches that of the target. The token $x_j$ immediately following that earlier history is the corresponding copy candidate. If no such $j$ exists within the window, the target is classified as having no repeat. A predictor with context length $T$ conditions on $x_{t-T},\dots,x_{t-1}$ and can access both the earlier history $h_j$ and its continuation $x_j$ exactly when $j-16 \ge t-T$, equivalently $\ell_t+16\le T$; we then say that the repeat is \emph{in view}. Because $j<t$, $x_j$ is always a past token, so the target $x_t$ is never exposed. Lag bins are defined as 1 to 32, 33 to 80, 81 to 176, and 177 to 320, such that each bin becomes fully visible at one tested context length (48, 96, 192, and 336 tokens, respectively) and is entirely out of view below that length. Histories consisting of a single held pitch are treated separately because they match at nearly every adjacent position. For the latent predictor, each target step spans $x_t,\dots,x_{t+5}$ and is predicted only from earlier latent steps. We therefore use the history $h_t$ preceding its first token and require the nearest matching position to satisfy $j\le t-6$, ensuring that the candidate continuation $x_j,\dots,x_{j+5}$ ends before the target step begins. Sixteen tokens correspond to one 4/4 bar on the Nottingham grid and 2 seconds on the MAESTRO grid; we repeat the analysis with 8- and 32-token histories.

% ============================================================
\section{Results}
\label{sec:results}
% ============================================================

\subsection{The context sweep}
\label{sec:sweep}

\begin{figure}[t]
\centering
% generated by scripts/make_paper_tables.py; do not edit
\definecolor{clat}{HTML}{2A78D6}
\definecolor{ctok}{HTML}{EB6834}
\definecolor{cink}{HTML}{0B0B0B}
\definecolor{cink2}{HTML}{52514E}
\definecolor{cmuted}{HTML}{898781}
\definecolor{cgrid}{HTML}{E1E0D9}
\begin{tikzpicture}
\begin{axis}[name=p0,  width=0.9\columnwidth, height=4.1cm, xmode=log, log basis x=2, clip=false,
  xmin=4.5, xmax=640, ymin=0.2, ymax=1.15, xtick={6,18,48,96,192,336}, xticklabels={6,18,48,96,192,336},
  ylabel={NLL / NLL at 6 tokens}, 
  title={(a) Nottingham}, title style={font=\footnotesize, text=cink, yshift=-1ex}, legend to name=sweeplegend, legend columns=-1, legend style={font=\footnotesize, draw=none, text=cink, /tikz/every even column/.append style={column sep=0.35cm}},
  grid=major, grid style={cgrid, line width=0.3pt}, axis line style={cmuted}, tick style={cmuted}, tick label style={font=\footnotesize, text=cink2}, label style={font=\footnotesize, text=cink}]
\draw[cmuted, dotted, line width=0.6pt] (axis cs:96,0.2) -- (axis cs:96,1.15);
\addplot[clat, line width=1.2pt, mark=*, mark size=1.7pt, error bars/.cd, y dir=both, y explicit] coordinates {(6,1.0000) +- (0,0.0029) (18,0.9826) +- (0,0.0021) (48,0.9738) +- (0,0.0057) (96,0.9052) +- (0,0.0051) (192,0.6662) +- (0,0.0175) (336,0.5430) +- (0,0.0109)};
\addlegendentry{latent}
\addplot[ctok, line width=1.2pt, densely dashed, mark=square*, mark size=1.6pt, mark options={solid}, error bars/.cd, y dir=both, y explicit] coordinates {(6,1.0000) +- (0,0.0108) (18,1.0352) +- (0,0.0147) (48,1.0991) +- (0,0.0645) (96,0.9158) +- (0,0.2343) (192,0.4075) +- (0,0.0056) (336,0.2913) +- (0,0.0044)};
\addlegendentry{token}
\addplot[only marks, clat!55!white, mark=o, mark size=1.0pt, forget plot] coordinates {(5.66,0.9982) (5.66,1.0033) (5.66,0.9985) (16.98,0.9804) (16.98,0.9846) (16.98,0.9827) (45.28,0.9734) (45.28,0.9797) (45.28,0.9683) (90.57,0.9025) (90.57,0.9110) (90.57,0.9019) (181.13,0.6610) (181.13,0.6857) (181.13,0.6519) (316.98,0.5506) (316.98,0.5479) (316.98,0.5305)};
\addplot[only marks, ctok!55!white, mark=square, mark size=1.0pt, forget plot] coordinates {(6.36,0.9903) (6.36,1.0116) (6.36,0.9981) (19.08,1.0410) (19.08,1.0461) (19.08,1.0184) (50.88,1.1582) (50.88,1.0304) (50.88,1.1089) (101.76,1.0933) (101.76,0.6502) (101.76,1.0038) (203.52,0.4095) (203.52,0.4119) (203.52,0.4012) (356.16,0.2918) (356.16,0.2866) (356.16,0.2954)};
\node[font=\footnotesize, text=clat, anchor=west] at (axis cs:370,0.543) {latent $-$46\%};
\node[font=\footnotesize, text=ctok, anchor=west] at (axis cs:370,0.291) {token $-$71\%};
\end{axis}
\begin{axis}[name=p1, at={(p0.below south west)}, anchor=north west, yshift=-0.9cm, width=0.9\columnwidth, height=4.1cm, xmode=log, log basis x=2, clip=false,
  xmin=4.5, xmax=640, ymin=0.2, ymax=1.15, xtick={6,18,48,96,192,336}, xticklabels={6,18,48,96,192,336},
  ylabel={NLL / NLL at 6 tokens}, 
  title={(b) O'Neill's}, title style={font=\footnotesize, text=cink, yshift=-1ex}, 
  grid=major, grid style={cgrid, line width=0.3pt}, axis line style={cmuted}, tick style={cmuted}, tick label style={font=\footnotesize, text=cink2}, label style={font=\footnotesize, text=cink}]
\draw[cmuted, dotted, line width=0.6pt] (axis cs:96,0.2) -- (axis cs:96,1.15);
\addplot[clat, line width=1.2pt, mark=*, mark size=1.7pt, error bars/.cd, y dir=both, y explicit] coordinates {(6,1.0000) +- (0,0.0063) (18,0.9670) +- (0,0.0039) (48,0.9126) +- (0,0.0061) (96,0.8637) +- (0,0.0102) (192,0.6861) +- (0,0.0026) (336,0.4951) +- (0,0.0084)};

\addplot[ctok, line width=1.2pt, densely dashed, mark=square*, mark size=1.6pt, mark options={solid}, error bars/.cd, y dir=both, y explicit] coordinates {(6,1.0000) +- (0,0.0083) (18,0.9121) +- (0,0.0113) (48,0.8566) +- (0,0.0098) (96,0.6853) +- (0,0.0088) (192,0.4673) +- (0,0.0071) (336,0.2935) +- (0,0.0067)};

\addplot[only marks, clat!55!white, mark=o, mark size=1.0pt, forget plot] coordinates {(5.66,0.9979) (5.66,0.9950) (5.66,1.0071) (16.98,0.9692) (16.98,0.9626) (16.98,0.9694) (45.28,0.9068) (45.28,0.9189) (45.28,0.9120) (90.57,0.8521) (90.57,0.8673) (90.57,0.8716) (181.13,0.6852) (181.13,0.6840) (181.13,0.6890) (316.98,0.4857) (316.98,0.5020) (316.98,0.4976)};
\addplot[only marks, ctok!55!white, mark=square, mark size=1.0pt, forget plot] coordinates {(6.36,1.0042) (6.36,0.9905) (6.36,1.0053) (19.08,0.9214) (19.08,0.8996) (19.08,0.9154) (50.88,0.8651) (50.88,0.8459) (50.88,0.8590) (101.76,0.6774) (101.76,0.6838) (101.76,0.6948) (203.52,0.4703) (203.52,0.4592) (203.52,0.4724) (356.16,0.3006) (356.16,0.2872) (356.16,0.2927)};
\node[font=\footnotesize, text=clat, anchor=west] at (axis cs:370,0.495) {latent $-$50\%};
\node[font=\footnotesize, text=ctok, anchor=west] at (axis cs:370,0.294) {token $-$71\%};
\end{axis}
\begin{axis}[name=p2, at={(p1.below south west)}, anchor=north west, yshift=-0.9cm, width=0.9\columnwidth, height=4.1cm, xmode=log, log basis x=2, clip=false,
  xmin=4.5, xmax=640, ymin=0.2, ymax=1.15, xtick={6,18,48,96,192,336}, xticklabels={6,18,48,96,192,336},
  ylabel={NLL / NLL at 6 tokens}, 
  title={(c) MusicNet}, title style={font=\footnotesize, text=cink, yshift=-1ex}, 
  grid=major, grid style={cgrid, line width=0.3pt}, axis line style={cmuted}, tick style={cmuted}, tick label style={font=\footnotesize, text=cink2}, label style={font=\footnotesize, text=cink}]
\draw[cmuted, dotted, line width=0.6pt] (axis cs:96,0.2) -- (axis cs:96,1.15);

\addplot[ctok, line width=1.2pt, densely dashed, mark=square*, mark size=1.6pt, mark options={solid}, error bars/.cd, y dir=both, y explicit] coordinates {(6,1.0000) +- (0,0.0265) (18,0.9420) +- (0,0.0013) (48,0.8597) +- (0,0.0015) (96,0.8279) +- (0,0.0012) (192,0.8163) +- (0,0.0035) (336,0.8229) +- (0,0.0018)};

\addplot[only marks, ctok!55!white, mark=square, mark size=1.0pt, forget plot] coordinates {(6.36,1.0289) (6.36,0.9942) (6.36,0.9769) (19.08,0.9405) (19.08,0.9426) (19.08,0.9429) (50.88,0.8611) (50.88,0.8582) (50.88,0.8597) (101.76,0.8267) (101.76,0.8291) (101.76,0.8280) (203.52,0.8127) (203.52,0.8198) (203.52,0.8165) (356.16,0.8249) (356.16,0.8217) (356.16,0.8220)};

\node[font=\footnotesize, text=ctok, anchor=west] at (axis cs:370,0.823) {token $-$18\%};
\end{axis}
\begin{axis}[name=p3, at={(p2.below south west)}, anchor=north west, yshift=-0.9cm, width=0.9\columnwidth, height=4.1cm, xmode=log, log basis x=2, clip=false,
  xmin=4.5, xmax=640, ymin=0.2, ymax=1.15, xtick={6,18,48,96,192,336}, xticklabels={6,18,48,96,192,336},
  ylabel={NLL / NLL at 6 tokens}, xlabel={Context (tokens)},
  title={(d) MAESTRO}, title style={font=\footnotesize, text=cink, yshift=-1ex}, 
  grid=major, grid style={cgrid, line width=0.3pt}, axis line style={cmuted}, tick style={cmuted}, tick label style={font=\footnotesize, text=cink2}, label style={font=\footnotesize, text=cink}]
\draw[cmuted, dotted, line width=0.6pt] (axis cs:96,0.2) -- (axis cs:96,1.15);
\addplot[clat, line width=1.2pt, mark=*, mark size=1.7pt, error bars/.cd, y dir=both, y explicit] coordinates {(6,1.0000) +- (0,0.0005) (18,0.9937) +- (0,0.0000) (48,0.9910) +- (0,0.0005) (96,0.9890) +- (0,0.0004) (192,0.9884) +- (0,0.0001) (336,0.9885) +- (0,0.0005)};

\addplot[ctok, line width=1.2pt, densely dashed, mark=square*, mark size=1.6pt, mark options={solid}, error bars/.cd, y dir=both, y explicit] coordinates {(6,1.0000) +- (0,0.0007) (18,0.9677) +- (0,0.0007) (48,0.9434) +- (0,0.0018) (96,0.9262) +- (0,0.0026) (192,0.9178) +- (0,0.0017) (336,0.9098) +- (0,0.0016)};

\addplot[only marks, clat!55!white, mark=o, mark size=1.0pt, forget plot] coordinates {(5.66,0.9995) (5.66,1.0005) (5.66,1.0000) (16.98,0.9937) (16.98,0.9937) (16.98,0.9937) (45.28,0.9905) (45.28,0.9915) (45.28,0.9910) (90.57,0.9885) (90.57,0.9890) (90.57,0.9894) (181.13,0.9883) (181.13,0.9885) (181.13,0.9884) (316.98,0.9881) (316.98,0.9884) (316.98,0.9890)};
\addplot[only marks, ctok!55!white, mark=square, mark size=1.0pt, forget plot] coordinates {(6.36,0.9992) (6.36,1.0006) (6.36,1.0002) (19.08,0.9676) (19.08,0.9670) (19.08,0.9684) (50.88,0.9424) (50.88,0.9424) (50.88,0.9455) (101.76,0.9245) (101.76,0.9292) (101.76,0.9249) (203.52,0.9170) (203.52,0.9167) (203.52,0.9197) (356.16,0.9093) (356.16,0.9085) (356.16,0.9116)};
\node[font=\footnotesize, text=clat, anchor=west] at (axis cs:370,1.029) {latent $-$1\%};
\node[font=\footnotesize, text=ctok, anchor=west] at (axis cs:370,0.869) {token $-$9\%};
\end{axis}
\end{tikzpicture}
\par\vspace{2pt}\ref{sweeplegend}
\caption{Test NLL relative to its value at 6 tokens of context for (a) Nottingham, (b) O'Neill's, (c) MusicNet and (d) MAESTRO. Lines are the mean over three seeds with the seed SD as error bars; the small hollow marks beside them are the individual seeds. The label at the end of each curve is its drop from 6 to 336 tokens. The context axis is logarithmic, and the dotted line marks 96 tokens.}
\label{fig:sweep}
\end{figure}
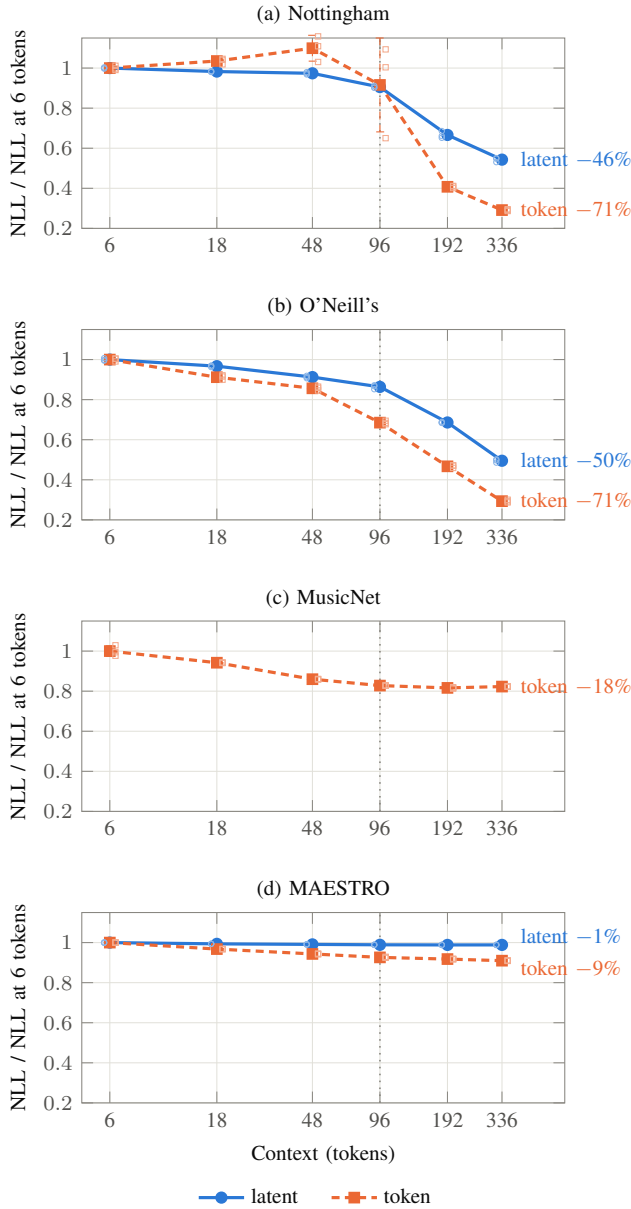

Table~\ref{tab:sweep} and Fig.~\ref{fig:sweep} summarize the context-length sweep.

\begin{table*}[t]
\centering
\caption{Test NLL against context length, mean and SD over three seeds. Context 0 is the unigram distribution (tokens) and the per-bit marginal (latents), each from one run. Lower is better. Nottingham and O'Neill's use transposition of the training split. MAESTRO tokens: 89-token data. MusicNet has token predictors only. Section~\ref{sec:stability} reports ten more token runs at 48, 96 and 192 tokens with a fixed training budget.}
\label{tab:sweep}
% generated by scripts/make_paper_tables.py; do not edit
\begin{tabular}{rcc@{\hspace{1.4em}}cc@{\hspace{1.4em}}c@{\hspace{1.4em}}cc}
\toprule
 & \multicolumn{2}{c}{\textbf{Nottingham}} & \multicolumn{2}{c}{\textbf{O'Neill's}} & \multicolumn{1}{c}{\textbf{MusicNet}} & \multicolumn{2}{c}{\textbf{MAESTRO}} \\
\cmidrule(lr){2-3}\cmidrule(lr){4-5}\cmidrule(lr){6-6}\cmidrule(lr){7-8}
\textbf{Context} & Latent & Token & Latent & Token & Token & Latent & Token \\
\textbf{(tokens)} & (bits per bit) & (bits per token) & (bits per bit) & (bits per token) & (bits per token) & (bits per bit) & (bits per token) \\
\midrule
0 & 0.958 & 4.366 & 0.931 & 4.383 & 5.295 & 0.978 & 5.319 \\
6 & 0.857 $\pm$ 0.002 & 1.391 $\pm$ 0.015 & 0.845 $\pm$ 0.005 & 1.692 $\pm$ 0.014 & 2.512 $\pm$ 0.066 & 0.887 $\pm$ 0.0004 & 3.283 $\pm$ 0.002 \\
18 & 0.842 $\pm$ 0.002 & 1.440 $\pm$ 0.021 & 0.818 $\pm$ 0.003 & 1.543 $\pm$ 0.019 & 2.366 $\pm$ 0.003 & 0.881 $\pm$ 0.00003 & 3.177 $\pm$ 0.002 \\
48 & 0.834 $\pm$ 0.005 & 1.529 $\pm$ 0.090 & 0.771 $\pm$ 0.005 & 1.449 $\pm$ 0.017 & 2.159 $\pm$ 0.004 & 0.879 $\pm$ 0.0004 & 3.097 $\pm$ 0.006 \\
96 & 0.775 $\pm$ 0.004 & 1.274 $\pm$ 0.326 & 0.730 $\pm$ 0.009 & 1.160 $\pm$ 0.015 & 2.080 $\pm$ 0.003 & 0.877 $\pm$ 0.0004 & 3.040 $\pm$ 0.009 \\
192 & 0.571 $\pm$ 0.015 & 0.567 $\pm$ 0.008 & 0.580 $\pm$ 0.002 & 0.791 $\pm$ 0.012 & \textbf{2.050 $\pm$ 0.009} & \textbf{0.877 $\pm$ 0.00010} & 3.013 $\pm$ 0.005 \\
336 & \textbf{0.465 $\pm$ 0.009} & \textbf{0.405 $\pm$ 0.006} & \textbf{0.419 $\pm$ 0.007} & \textbf{0.497 $\pm$ 0.011} & 2.067 $\pm$ 0.004 & 0.877 $\pm$ 0.0004 & \textbf{2.987 $\pm$ 0.005} \\
\bottomrule
\end{tabular}

\end{table*}

On Nottingham, the latent predictor improves monotonically across the tested context lengths, with modest gains up to 48 tokens and a substantially steeper reduction after 96: from \R{nott/lat/6} bits per bit at 6 tokens to \R{nott/lat/96} at 96, \R{nott/lat/192} at 192, and \R{nott/lat/336} at 336. The token predictor initially degrades from 6 to 48 tokens (\R{nott/tok/6}, \R{nott/tok/18}, and \R{nott/tok/48} bits per token), exhibits substantial seed variation at 96 tokens, and then improves sharply to \R{nott/tok/192} at 192 tokens and \R{nott/tok/336} at 336, with a seed SD of \R{nott/tok/336/sd}. From 6 to 336 tokens, the reduction is \R{nott/boot/t/6-336} bits per token for the token predictor (95\% interval \R{nott/boot/t/6-336/lo} to \R{nott/boot/t/6-336/hi}) and \R{nott/boot/l/6-336} bits per bit for the latent predictor (\R{nott/boot/l/6-336/lo} to \R{nott/boot/l/6-336/hi}). The initial degradation of the token predictor is also supported by the paired intervals: defining a reduction as previous-context NLL minus longer-context NLL, the changes are $\R{nott/boot/t/6-18}$ bits from 6 to 18 tokens ($\R{nott/boot/t/6-18/lo}$ to $\R{nott/boot/t/6-18/hi}$) and $\R{nott/boot/t/18-48}$ bits from 18 to 48 ($\R{nott/boot/t/18-48/lo}$ to $\R{nott/boot/t/18-48/hi}$); both intervals exclude zero.

On MAESTRO, most of the improvement occurs at shorter context lengths. The token predictor decreases from \R{maes/tok/6} bits per token at 6 tokens to \R{maes/tok/96} at 96 and \R{maes/tok/336} at 336, reaching \R{maes/reached/t/96}\% of its total 6-to-336-token reduction by 96 tokens. The remaining 96-to-336-token reduction is small but consistently positive: \R{maes/boot/t/96-336} bits (\R{maes/boot/t/96-336/lo} to \R{maes/boot/t/96-336/hi}). The latent predictor decreases from \R{maes/lat/6} to \R{maes/lat/48} bits per bit by 48 tokens and changes by only \R{maes/boot/l/96-336} bits per bit from 96 to 336 tokens; the corresponding interval, $\R{maes/boot/l/96-336/lo}$ to \R{maes/boot/l/96-336/hi}, includes zero. No MAESTRO token result has a seed SD above \R{maes/tok/maxsd}. Because the reported token results use the unclamped 89-token representation, their absolute bits-per-token values are not directly comparable with the 48-token runs; however, the context-length trend is similar. In the clamped runs, the 6-to-336-token reduction is \R{maes48/boot/t/6-336} bits, \R{maes48/reached/t/96}\% of which is attained by 96 tokens, and the remaining 96-to-336-token reduction is \R{maes48/boot/t/96-336} bits. All of these statements are restricted to the tested range; we do not evaluate contexts beyond 336 tokens.

\subsection{Where the gain comes from}
\label{sec:lag}

Table~\ref{tab:lag} reports the distribution of targets across repeat-lag bins. On Nottingham, \R{nott/share/t/repeat}\% of target tokens have an earlier occurrence of their 16-token history within 320 tokens, with the largest mass at lags of 81 to 176 tokens, corresponding to approximately 5 to 11 bars in 4/4. For \R{nott/copy/t/all}\% of these targets, the token following the earlier occurrence equals the current target, making exact copying a valid prediction. In MAESTRO, by contrast, \R{maes/share/t/none}\% of target tokens have no such repeat.

\begin{table*}[t]
\centering
\caption{Repeat lag of the test targets: the share of targets whose 16-token history last occurred at each lag, or not within 320 tokens (none), or is one held pitch (held). A repeat is in view for context $T$ when lag $+$ 16 $\le T$, and each bin comes into view at the context shown. Latent steps use the history before the step's first token. The last rows describe the Nottingham and O'Neill's repeats: how often the next token, or the next 6 tokens, equal those that followed the earlier occurrence, and how often the earlier occurrence starts on the latent step grid, where the same 6 tokens have the same code.}
\label{tab:lag}
% generated by scripts/make_paper_tables.py; do not edit
\begin{tabular}{lcccccc}
\toprule
 & \multicolumn{4}{c}{\textbf{Repeat lag (tokens back)}} & & \\
\cmidrule(lr){2-5}
\textbf{Share of targets} & 1-32 & 33-80 & 81-176 & 177-320 & \textbf{None} & \textbf{Held} \\
\emph{in view from context} & \emph{48} & \emph{96} & \emph{192} & \emph{336} & & \\
\midrule
Nottingham, target tokens & 1.8\% & 14.7\% & 36.6\% & 17.9\% & 28.3\% & 0.7\% \\
Nottingham, target latent steps & 1.9\% & 15.4\% & 36.4\% & 17.7\% & 27.5\% & 1.2\% \\
O'Neill's, target tokens & 0.3\% & 9.6\% & 33.3\% & 30.1\% & 26.7\% & 0.0\% \\
O'Neill's, target latent steps & 0.4\% & 11.1\% & 33.8\% & 29.1\% & 25.6\% & 0.0\% \\
MusicNet, target tokens & 3.1\% & 2.8\% & 2.4\% & 2.2\% & 86.4\% & 3.1\% \\
MAESTRO, target tokens & 0.1\% & 0.2\% & 0.2\% & 0.2\% & 96.9\% & 2.3\% \\
MAESTRO, target latent steps & 0.1\% & 0.2\% & 0.2\% & 0.3\% & 96.8\% & 2.4\% \\
\midrule
\emph{Properties of the repeats} & & & & & & \\
Nottingham, next token repeats & 82\% & 94\% & 99\% & 99\% & n/a & n/a \\
Nottingham, next 6 tokens repeat & 35\% & 70\% & 92\% & 94\% & n/a & n/a \\
Nottingham, occurrence on the step grid & 24\% & 22\% & 31\% & 54\% & n/a & n/a \\
O'Neill's, next token repeats & 62\% & 90\% & 98\% & 99\% & n/a & n/a \\
O'Neill's, next 6 tokens repeat & 12\% & 53\% & 88\% & 95\% & n/a & n/a \\
O'Neill's, occurrence on the step grid & 54\% & 82\% & 67\% & 79\% & n/a & n/a \\
\bottomrule
\end{tabular}

\end{table*}

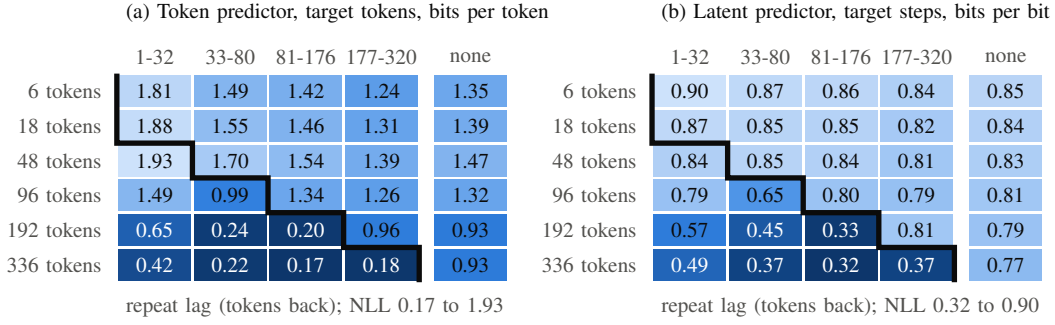
\begin{figure*}[t]
\centering
% generated by scripts/make_paper_tables.py; do not edit
\definecolor{clat}{HTML}{2A78D6}
\definecolor{ctok}{HTML}{EB6834}
\definecolor{cink}{HTML}{0B0B0B}
\definecolor{cink2}{HTML}{52514E}
\definecolor{cmuted}{HTML}{898781}
\definecolor{cgrid}{HTML}{E1E0D9}
\begin{tikzpicture}[font=\footnotesize]
\node[anchor=south west, text=cink, font=\footnotesize] at (1.35,3.31) {(a) Token predictor, target tokens, bits per token};
\node[text=cink2] at (1.85,2.98) {1-32};
\node[text=cink2] at (2.85,2.98) {33-80};
\node[text=cink2] at (3.85,2.98) {81-176};
\node[text=cink2] at (4.85,2.98) {177-320};
\node[text=cink2] at (6.03,2.98) {none};
\node[anchor=east, text=cink2] at (1.27,2.53) {6 tokens};
\fill[fill={rgb,255:red,187;green,214;blue,247}] (1.37,2.32) rectangle (2.33,2.74);
\node[text=cink] at (1.85,2.53) {1.81};
\fill[fill={rgb,255:red,132;green,181;blue,239}] (2.37,2.32) rectangle (3.33,2.74);
\node[text=cink] at (2.85,2.53) {1.49};
\fill[fill={rgb,255:red,121;green,174;blue,237}] (3.37,2.32) rectangle (4.33,2.74);
\node[text=cink] at (3.85,2.53) {1.42};
\fill[fill={rgb,255:red,92;green,156;blue,232}] (4.37,2.32) rectangle (5.33,2.74);
\node[text=cink] at (4.85,2.53) {1.24};
\fill[fill={rgb,255:red,109;green,167;blue,236}] (5.55,2.32) rectangle (6.51,2.74);
\node[text=cink] at (6.03,2.53) {1.35};
\node[anchor=east, text=cink2] at (1.27,2.07) {18 tokens};
\fill[fill={rgb,255:red,196;green,220;blue,249}] (1.37,1.86) rectangle (2.33,2.28);
\node[text=cink] at (1.85,2.07) {1.88};
\fill[fill={rgb,255:red,144;green,188;blue,241}] (2.37,1.86) rectangle (3.33,2.28);
\node[text=cink] at (2.85,2.07) {1.55};
\fill[fill={rgb,255:red,128;green,178;blue,238}] (3.37,1.86) rectangle (4.33,2.28);
\node[text=cink] at (3.85,2.07) {1.46};
\fill[fill={rgb,255:red,102;green,163;blue,235}] (4.37,1.86) rectangle (5.33,2.28);
\node[text=cink] at (4.85,2.07) {1.31};
\fill[fill={rgb,255:red,116;green,171;blue,237}] (5.55,1.86) rectangle (6.51,2.28);
\node[text=cink] at (6.03,2.07) {1.39};
\node[anchor=east, text=cink2] at (1.27,1.61) {48 tokens};
\fill[fill={rgb,255:red,205;green,226;blue,251}] (1.37,1.40) rectangle (2.33,1.82);
\node[text=cink] at (1.85,1.61) {1.93};
\fill[fill={rgb,255:red,167;green,202;blue,245}] (2.37,1.40) rectangle (3.33,1.82);
\node[text=cink] at (2.85,1.61) {1.70};
\fill[fill={rgb,255:red,142;green,187;blue,241}] (3.37,1.40) rectangle (4.33,1.82);
\node[text=cink] at (3.85,1.61) {1.54};
\fill[fill={rgb,255:red,116;green,171;blue,237}] (4.37,1.40) rectangle (5.33,1.82);
\node[text=cink] at (4.85,1.61) {1.39};
\fill[fill={rgb,255:red,130;green,179;blue,238}] (5.55,1.40) rectangle (6.51,1.82);
\node[text=cink] at (6.03,1.61) {1.47};
\node[anchor=east, text=cink2] at (1.27,1.15) {96 tokens};
\fill[fill={rgb,255:red,134;green,182;blue,239}] (1.37,0.94) rectangle (2.33,1.36);
\node[text=cink] at (1.85,1.15) {1.49};
\fill[fill={rgb,255:red,50;green,128;blue,222}] (2.37,0.94) rectangle (3.33,1.36);
\node[text=cink] at (2.85,1.15) {0.99};
\fill[fill={rgb,255:red,108;green,166;blue,236}] (3.37,0.94) rectangle (4.33,1.36);
\node[text=cink] at (3.85,1.15) {1.34};
\fill[fill={rgb,255:red,94;green,158;blue,233}] (4.37,0.94) rectangle (5.33,1.36);
\node[text=cink] at (4.85,1.15) {1.26};
\fill[fill={rgb,255:red,104;green,164;blue,235}] (5.55,0.94) rectangle (6.51,1.36);
\node[text=cink] at (6.03,1.15) {1.32};
\node[anchor=east, text=cink2] at (1.27,0.69) {192 tokens};
\fill[fill={rgb,255:red,31;green,96;blue,177}] (1.37,0.48) rectangle (2.33,0.90);
\node[text=white] at (1.85,0.69) {0.65};
\fill[fill={rgb,255:red,14;green,60;blue,118}] (2.37,0.48) rectangle (3.33,0.90);
\node[text=white] at (2.85,0.69) {0.24};
\fill[fill={rgb,255:red,14;green,57;blue,112}] (3.37,0.48) rectangle (4.33,0.90);
\node[text=white] at (3.85,0.69) {0.20};
\fill[fill={rgb,255:red,47;green,125;blue,219}] (4.37,0.48) rectangle (5.33,0.90);
\node[text=cink] at (4.85,0.69) {0.96};
\fill[fill={rgb,255:red,45;green,123;blue,217}] (5.55,0.48) rectangle (6.51,0.90);
\node[text=cink] at (6.03,0.69) {0.93};
\node[anchor=east, text=cink2] at (1.27,0.23) {336 tokens};
\fill[fill={rgb,255:red,22;green,75;blue,143}] (1.37,0.02) rectangle (2.33,0.44);
\node[text=white] at (1.85,0.23) {0.42};
\fill[fill={rgb,255:red,14;green,58;blue,113}] (2.37,0.02) rectangle (3.33,0.44);
\node[text=white] at (2.85,0.23) {0.22};
\fill[fill={rgb,255:red,13;green,54;blue,107}] (3.37,0.02) rectangle (4.33,0.44);
\node[text=white] at (3.85,0.23) {0.17};
\fill[fill={rgb,255:red,13;green,55;blue,108}] (4.37,0.02) rectangle (5.33,0.44);
\node[text=white] at (4.85,0.23) {0.18};
\fill[fill={rgb,255:red,45;green,123;blue,217}] (5.55,0.02) rectangle (6.51,0.44);
\node[text=cink] at (6.03,0.23) {0.93};
\draw[cink, line width=2pt] (1.35,2.76) -- (1.35,2.30) -- (1.35,2.30) -- (1.35,1.84) -- (2.35,1.84) -- (2.35,1.38) -- (3.35,1.38) -- (3.35,0.92) -- (4.35,0.92) -- (4.35,0.46) -- (5.35,0.46) -- (5.35,0.00);
\node[anchor=north west, text=cink2] at (1.35,-0.08) {repeat lag (tokens back); NLL 0.17 to 1.93};
\node[anchor=south west, text=cink, font=\footnotesize] at (8.43,3.31) {(b) Latent predictor, target steps, bits per bit};
\node[text=cink2] at (8.93,2.98) {1-32};
\node[text=cink2] at (9.93,2.98) {33-80};
\node[text=cink2] at (10.93,2.98) {81-176};
\node[text=cink2] at (11.93,2.98) {177-320};
\node[text=cink2] at (13.11,2.98) {none};
\node[anchor=east, text=cink2] at (8.35,2.53) {6 tokens};
\fill[fill={rgb,255:red,205;green,226;blue,251}] (8.45,2.32) rectangle (9.41,2.74);
\node[text=cink] at (8.93,2.53) {0.90};
\fill[fill={rgb,255:red,191;green,217;blue,248}] (9.45,2.32) rectangle (10.41,2.74);
\node[text=cink] at (9.93,2.53) {0.87};
\fill[fill={rgb,255:red,188;green,214;blue,247}] (10.45,2.32) rectangle (11.41,2.74);
\node[text=cink] at (10.93,2.53) {0.86};
\fill[fill={rgb,255:red,175;green,207;blue,245}] (11.45,2.32) rectangle (12.41,2.74);
\node[text=cink] at (11.93,2.53) {0.84};
\fill[fill={rgb,255:red,185;green,212;blue,246}] (12.63,2.32) rectangle (13.59,2.74);
\node[text=cink] at (13.11,2.53) {0.85};
\node[anchor=east, text=cink2] at (8.35,2.07) {18 tokens};
\fill[fill={rgb,255:red,191;green,216;blue,248}] (8.45,1.86) rectangle (9.41,2.28);
\node[text=cink] at (8.93,2.07) {0.87};
\fill[fill={rgb,255:red,184;green,212;blue,246}] (9.45,1.86) rectangle (10.41,2.28);
\node[text=cink] at (9.93,2.07) {0.85};
\fill[fill={rgb,255:red,181;green,210;blue,246}] (10.45,1.86) rectangle (11.41,2.28);
\node[text=cink] at (10.93,2.07) {0.85};
\fill[fill={rgb,255:red,166;green,201;blue,245}] (11.45,1.86) rectangle (12.41,2.28);
\node[text=cink] at (11.93,2.07) {0.82};
\fill[fill={rgb,255:red,178;green,208;blue,246}] (12.63,1.86) rectangle (13.59,2.28);
\node[text=cink] at (13.11,2.07) {0.84};
\node[anchor=east, text=cink2] at (8.35,1.61) {48 tokens};
\fill[fill={rgb,255:red,176;green,207;blue,245}] (8.45,1.40) rectangle (9.41,1.82);
\node[text=cink] at (8.93,1.61) {0.84};
\fill[fill={rgb,255:red,183;green,211;blue,246}] (9.45,1.40) rectangle (10.41,1.82);
\node[text=cink] at (9.93,1.61) {0.85};
\fill[fill={rgb,255:red,178;green,208;blue,246}] (10.45,1.40) rectangle (11.41,1.82);
\node[text=cink] at (10.93,1.61) {0.84};
\fill[fill={rgb,255:red,161;green,199;blue,244}] (11.45,1.40) rectangle (12.41,1.82);
\node[text=cink] at (11.93,1.61) {0.81};
\fill[fill={rgb,255:red,173;green,206;blue,245}] (12.63,1.40) rectangle (13.59,1.82);
\node[text=cink] at (13.11,1.61) {0.83};
\node[anchor=east, text=cink2] at (8.35,1.15) {96 tokens};
\fill[fill={rgb,255:red,152;green,193;blue,243}] (8.45,0.94) rectangle (9.41,1.36);
\node[text=cink] at (8.93,1.15) {0.79};
\fill[fill={rgb,255:red,79;green,149;blue,231}] (9.45,0.94) rectangle (10.41,1.36);
\node[text=cink] at (9.93,1.15) {0.65};
\fill[fill={rgb,255:red,155;green,195;blue,243}] (10.45,0.94) rectangle (11.41,1.36);
\node[text=cink] at (10.93,1.15) {0.80};
\fill[fill={rgb,255:red,150;green,192;blue,242}] (11.45,0.94) rectangle (12.41,1.36);
\node[text=cink] at (11.93,1.15) {0.79};
\fill[fill={rgb,255:red,161;green,199;blue,244}] (12.63,0.94) rectangle (13.59,1.36);
\node[text=cink] at (13.11,1.15) {0.81};
\node[anchor=east, text=cink2] at (8.35,0.69) {192 tokens};
\fill[fill={rgb,255:red,46;green,124;blue,218}] (8.45,0.48) rectangle (9.41,0.90);
\node[text=cink] at (8.93,0.69) {0.57};
\fill[fill={rgb,255:red,27;green,88;blue,163}] (9.45,0.48) rectangle (10.41,0.90);
\node[text=white] at (9.93,0.69) {0.45};
\fill[fill={rgb,255:red,14;green,58;blue,114}] (10.45,0.48) rectangle (11.41,0.90);
\node[text=white] at (10.93,0.69) {0.33};
\fill[fill={rgb,255:red,161;green,199;blue,244}] (11.45,0.48) rectangle (12.41,0.90);
\node[text=cink] at (11.93,0.69) {0.81};
\fill[fill={rgb,255:red,154;green,194;blue,243}] (12.63,0.48) rectangle (13.59,0.90);
\node[text=cink] at (13.11,0.69) {0.79};
\node[anchor=east, text=cink2] at (8.35,0.23) {336 tokens};
\fill[fill={rgb,255:red,34;green,101;blue,184}] (8.45,0.02) rectangle (9.41,0.44);
\node[text=white] at (8.93,0.23) {0.49};
\fill[fill={rgb,255:red,16;green,67;blue,130}] (9.45,0.02) rectangle (10.41,0.44);
\node[text=white] at (9.93,0.23) {0.37};
\fill[fill={rgb,255:red,13;green,54;blue,107}] (10.45,0.02) rectangle (11.41,0.44);
\node[text=white] at (10.93,0.23) {0.32};
\fill[fill={rgb,255:red,16;green,67;blue,130}] (11.45,0.02) rectangle (12.41,0.44);
\node[text=white] at (11.93,0.23) {0.37};
\fill[fill={rgb,255:red,140;green,186;blue,240}] (12.63,0.02) rectangle (13.59,0.44);
\node[text=cink] at (13.11,0.23) {0.77};
\draw[cink, line width=2pt] (8.43,2.76) -- (8.43,2.30) -- (8.43,2.30) -- (8.43,1.84) -- (9.43,1.84) -- (9.43,1.38) -- (10.43,1.38) -- (10.43,0.92) -- (11.43,0.92) -- (11.43,0.46) -- (12.43,0.46) -- (12.43,0.00);
\node[anchor=north west, text=cink2] at (8.43,-0.08) {repeat lag (tokens back); NLL 0.32 to 0.90};
\end{tikzpicture}
\caption{Test NLL on Nottingham by repeat lag (columns) and context length (rows), mean over three seeds, for (a) the target tokens of the token predictor and (b) the target latent steps of the latent predictor, each on its own colour scale. Darker cells have lower NLL. Left of the black line the repeat is in view (lag $+$ 16 $\le$ context); right of it, it is not. The bins are cut at these limits.}
\label{fig:lag}
\end{figure*}

Fig.~\ref{fig:lag} stratifies Nottingham test NLL by repeat lag. Improvement appears when the earlier occurrence becomes accessible to the model, rather than before it enters the receptive field. At 6 tokens, the token predictor obtains between \R{nott/lag/t/6/177-320} and \R{nott/lag/t/6/1-32} bits per token across repeat bins. At 192 tokens, NLL decreases to \R{nott/lag/t/192/81-176} bits for repeats 81 to 176 tokens back and \R{nott/lag/t/192/33-80} for repeats 33 to 80 tokens back, both of which are in view, while repeats 177 to 320 tokens back remain out of view and score \R{nott/lag/t/192/177-320}; targets without repeats score \R{nott/lag/t/192/none}. When the context expands to 336 tokens, the 177 to 320 bin enters view and its NLL decreases to \R{nott/lag/t/336/177-320}. The latent predictor exhibits the same qualitative transition: at 192 tokens it scores \R{nott/lag/l/192/81-176} bits per bit on steps with repeats 81 to 176 tokens back and \R{nott/lag/l/192/177-320} on repeats 177 to 320 tokens back, the latter decreasing to \R{nott/lag/l/336/177-320} at 336 tokens. Only \R{nott/aligned/all}\% of latent-step repeats begin on the latent-step grid. Nevertheless, among copyable steps at 336 tokens, NLL is \R{nott/align/336/aligned} bits per bit when the earlier occurrence is grid-aligned and \R{nott/align/336/misaligned} when it is not, compared with \R{nott/lag/l/336/none} for steps with no repeat. Thus the latent predictor benefits even when corresponding repeated token spans are represented by different latent codes.

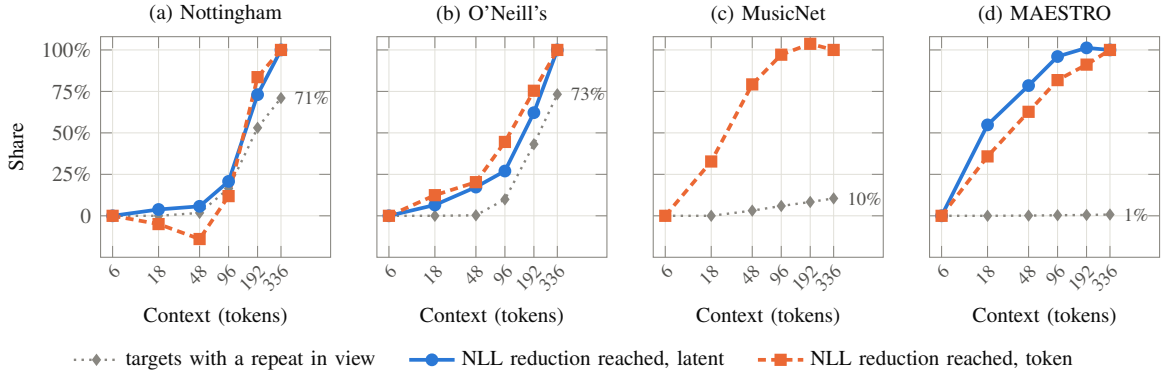
\begin{figure*}[t]
\centering
% generated by scripts/make_paper_tables.py; do not edit
\definecolor{clat}{HTML}{2A78D6}
\definecolor{ctok}{HTML}{EB6834}
\definecolor{cink}{HTML}{0B0B0B}
\definecolor{cink2}{HTML}{52514E}
\definecolor{cmuted}{HTML}{898781}
\definecolor{cgrid}{HTML}{E1E0D9}
\begin{tikzpicture}
\begin{axis}[name=q0,  width=0.255\textwidth, height=4.5cm, xmode=log, log basis x=2, clip=false,
  xmin=4.5, xmax=1100, ymin=-0.25, ymax=1.08, xtick={6,18,48,96,192,336}, xticklabels={6,18,48,96,192,336},
  ytick={0,0.25,0.5,0.75,1}, yticklabels={0,25\%,50\%,75\%,100\%},
  x tick label style={font=\scriptsize, rotate=45, anchor=north east, inner sep=1.5pt},
  ylabel={Share}, xlabel={Context (tokens)},
  title={(a) Nottingham}, title style={font=\footnotesize, text=cink, yshift=-1ex}, legend to name=coverlegend, legend columns=-1, legend style={font=\footnotesize, draw=none, text=cink, /tikz/every even column/.append style={column sep=0.35cm}},
  grid=major, grid style={cgrid, line width=0.3pt}, axis line style={cmuted}, tick style={cmuted}, tick label style={font=\footnotesize, text=cink2}, label style={font=\footnotesize, text=cink}]
\addplot[cmuted, line width=0.9pt, dotted, mark=diamond*, mark size=1.6pt, mark options={solid}] coordinates {(6,0.0000) (18,0.0000) (48,0.0175) (96,0.1646) (192,0.5305) (336,0.7100)};
\addlegendentry{targets with a repeat in view}
\addplot[clat, line width=1.4pt, mark=*, mark size=1.7pt] coordinates {(6,0.0000) (18,0.0382) (48,0.0573) (96,0.2075) (192,0.7304) (336,1.0000)};
\addlegendentry{NLL reduction reached, latent}
\addplot[ctok, line width=1.4pt, densely dashed, mark=square*, mark size=1.6pt, mark options={solid}] coordinates {(6,0.0000) (18,-0.0497) (48,-0.1399) (96,0.1188) (192,0.8360) (336,1.0000)};
\addlegendentry{NLL reduction reached, token}
\node[font=\scriptsize, text=cink2, anchor=west, inner sep=1pt, xshift=4pt] at (axis cs:336,0.710) {71\%};
\end{axis}
\begin{axis}[name=q1, at={(q0.right of south east)}, anchor=south west, xshift=0.6cm, width=0.255\textwidth, height=4.5cm, xmode=log, log basis x=2, clip=false,
  xmin=4.5, xmax=1100, ymin=-0.25, ymax=1.08, xtick={6,18,48,96,192,336}, xticklabels={6,18,48,96,192,336},
  ytick={0,0.25,0.5,0.75,1}, yticklabels={0,25\%,50\%,75\%,100\%},
  x tick label style={font=\scriptsize, rotate=45, anchor=north east, inner sep=1.5pt},
  yticklabels={,,}, xlabel={Context (tokens)},
  title={(b) O'Neill's}, title style={font=\footnotesize, text=cink, yshift=-1ex}, 
  grid=major, grid style={cgrid, line width=0.3pt}, axis line style={cmuted}, tick style={cmuted}, tick label style={font=\footnotesize, text=cink2}, label style={font=\footnotesize, text=cink}]
\addplot[cmuted, line width=0.9pt, dotted, mark=diamond*, mark size=1.6pt, mark options={solid}] coordinates {(6,0.0000) (18,0.0000) (48,0.0030) (96,0.0990) (192,0.4316) (336,0.7326)};

\addplot[clat, line width=1.4pt, mark=*, mark size=1.7pt] coordinates {(6,0.0000) (18,0.0653) (48,0.1732) (96,0.2700) (192,0.6218) (336,1.0000)};

\addplot[ctok, line width=1.4pt, densely dashed, mark=square*, mark size=1.6pt, mark options={solid}] coordinates {(6,0.0000) (18,0.1244) (48,0.2029) (96,0.4454) (192,0.7540) (336,1.0000)};

\node[font=\scriptsize, text=cink2, anchor=west, inner sep=1pt, xshift=4pt] at (axis cs:336,0.733) {73\%};
\end{axis}
\begin{axis}[name=q2, at={(q1.right of south east)}, anchor=south west, xshift=0.6cm, width=0.255\textwidth, height=4.5cm, xmode=log, log basis x=2, clip=false,
  xmin=4.5, xmax=1100, ymin=-0.25, ymax=1.08, xtick={6,18,48,96,192,336}, xticklabels={6,18,48,96,192,336},
  ytick={0,0.25,0.5,0.75,1}, yticklabels={0,25\%,50\%,75\%,100\%},
  x tick label style={font=\scriptsize, rotate=45, anchor=north east, inner sep=1.5pt},
  yticklabels={,,}, xlabel={Context (tokens)},
  title={(c) MusicNet}, title style={font=\footnotesize, text=cink, yshift=-1ex}, 
  grid=major, grid style={cgrid, line width=0.3pt}, axis line style={cmuted}, tick style={cmuted}, tick label style={font=\footnotesize, text=cink2}, label style={font=\footnotesize, text=cink}]
\addplot[cmuted, line width=0.9pt, dotted, mark=diamond*, mark size=1.6pt, mark options={solid}] coordinates {(6,0.0000) (18,0.0000) (48,0.0312) (96,0.0593) (192,0.0833) (336,0.1049)};

\addplot[ctok, line width=1.4pt, densely dashed, mark=square*, mark size=1.6pt, mark options={solid}] coordinates {(6,0.0000) (18,0.3275) (48,0.7923) (96,0.9714) (192,1.0370) (336,1.0000)};

\node[font=\scriptsize, text=cink2, anchor=west, inner sep=1pt, xshift=4pt] at (axis cs:336,0.105) {10\%};
\end{axis}
\begin{axis}[name=q3, at={(q2.right of south east)}, anchor=south west, xshift=0.6cm, width=0.255\textwidth, height=4.5cm, xmode=log, log basis x=2, clip=false,
  xmin=4.5, xmax=1100, ymin=-0.25, ymax=1.08, xtick={6,18,48,96,192,336}, xticklabels={6,18,48,96,192,336},
  ytick={0,0.25,0.5,0.75,1}, yticklabels={0,25\%,50\%,75\%,100\%},
  x tick label style={font=\scriptsize, rotate=45, anchor=north east, inner sep=1.5pt},
  yticklabels={,,}, xlabel={Context (tokens)},
  title={(d) MAESTRO}, title style={font=\footnotesize, text=cink, yshift=-1ex}, 
  grid=major, grid style={cgrid, line width=0.3pt}, axis line style={cmuted}, tick style={cmuted}, tick label style={font=\footnotesize, text=cink2}, label style={font=\footnotesize, text=cink}]
\addplot[cmuted, line width=0.9pt, dotted, mark=diamond*, mark size=1.6pt, mark options={solid}] coordinates {(6,0.0000) (18,0.0000) (48,0.0013) (96,0.0034) (192,0.0051) (336,0.0076)};

\addplot[clat, line width=1.4pt, mark=*, mark size=1.7pt] coordinates {(6,0.0000) (18,0.5482) (48,0.7851) (96,0.9603) (192,1.0125) (336,1.0000)};

\addplot[ctok, line width=1.4pt, densely dashed, mark=square*, mark size=1.6pt, mark options={solid}] coordinates {(6,0.0000) (18,0.3582) (48,0.6270) (96,0.8181) (192,0.9115) (336,1.0000)};

\node[font=\scriptsize, text=cink2, anchor=west, inner sep=1pt, xshift=4pt] at (axis cs:336,0.008) {1\%};
\end{axis}
\end{tikzpicture}
\par\vspace{2pt}\ref{coverlegend}
\caption{Share of target tokens whose repeat is in view (lag $+$ 16 $\le$ context, thin dotted line), and share of the NLL reduction from 6 to 336 tokens reached at each context (thick lines), for (a) Nottingham, (b) O'Neill's, (c) MusicNet and (d) MAESTRO. The label at 336 tokens is the repeat share there; the reduction reached is 100\% at 336 tokens by definition. A negative share means the NLL is above its value at 6 tokens.}
\label{fig:cover}
\end{figure*}

Most of the aggregate reduction is concentrated on repeated-history targets. Summed over targets, \R{nott/redshare/t/repeat}\% of the token predictor's 6-to-336-token NLL reduction comes from targets with a repeat in view at 336 tokens and \R{nott/redshare/t/none}\% from targets without one; the corresponding latent shares are \R{nott/redshare/l/repeat}\% and \R{nott/redshare/l/none}\%. Fig.~\ref{fig:cover} compares repeat availability with the fraction of the total NLL reduction attained at each context: the fraction of targets with a repeat in view increases from \R{nott/inview/48}\% at 48 tokens to \R{nott/inview/96}\% at 96, \R{nott/inview/192}\% at 192, and \R{nott/inview/336}\% at 336, while the token predictor reaches \R{nott/reached/t/96}\%, \R{nott/reached/t/192}\%, and 100\% of its eventual reduction at 96, 192, and 336 tokens. Using 8-token histories, \R{nott/h8/share}\% of targets have a repeat and account for \R{nott/h8/redshare/t}\% of the reduction; using 32-token histories, the corresponding values are \R{nott/h32/share}\% and \R{nott/h32/redshare/t}\%. On MAESTRO, only \R{maes/inview/336}\% of targets have a repeat in view at 336 tokens, and \R{maes/redshare/t/none}\% of the token predictor's reduction instead comes from targets without one. For the 89-token MAESTRO representation used by the token predictor, the exact repeated-target share is \R{maes/share/t/repeat2}\%, compared with \R{maes48/share/t/repeat}\% under clamping. Transposed repetitions do not account for the remaining Nottingham gain: for \R{nott/trans/share}\% of targets, the 16-token history occurs earlier only under transposition, and in \R{nott/trans/copy}\% of these cases the correspondingly transposed continuation equals the target. The 336-token token predictor nevertheless scores \R{nott/trans/t/336} bits on these targets, no better than the \R{nott/trans/other/336} bits obtained on other targets without an exact repeat, indicating no measurable copying across transposition in this setting.

\subsection{The predictors use the earlier occurrence}
\label{sec:copy}

\begin{table}[t]
\centering
\caption{Erase test on Nottingham and O'Neill's targets with a repeat in view, mean and SD over three seeds. Clean is the test NLL of these targets; erased overwrites every earlier occurrence of the 16-token history in the context, with what followed it; control overwrites as many spans of the same length elsewhere in the context. Latent values in bits per bit, token values in bits per token. The last column is the probability the token predictor gives to a new token placed after every earlier occurrence, before and after the swap.}
\label{tab:copy}
\resizebox{\columnwidth}{!}{% generated by scripts/make_paper_tables.py; do not edit
\begin{tabular}{lccccc}
\toprule
\textbf{Corpus, predictor, context} & \textbf{Targets} & \textbf{Clean} & \textbf{Erased} & \textbf{Control} & \textbf{Swapped token} \\
\midrule
Nottingham, token, 192 & 16{,}693 & 0.22 $\pm$ 0.005 & 1.10 $\pm$ 0.02 & 0.22 $\pm$ 0.004 & 0.5\% $\to$ 55\% \\
Nottingham, token, 336 & 22{,}454 & 0.19 $\pm$ 0.003 & 1.06 $\pm$ 0.03 & 0.19 $\pm$ 0.003 & 0.3\% $\to$ 62\% \\
Nottingham, latent, 192 & 2{,}783 & 0.37 $\pm$ 0.03 & 0.87 $\pm$ 0.01 & 0.38 $\pm$ 0.03 & n/a \\
Nottingham, latent, 336 & 3{,}747 & 0.35 $\pm$ 0.01 & 0.81 $\pm$ 0.004 & 0.35 $\pm$ 0.01 & n/a \\
O'Neill's, token, 192 & 31{,}068 & 0.37 $\pm$ 0.003 & 1.28 $\pm$ 0.01 & 0.38 $\pm$ 0.002 & 0.8\% $\to$ 51\% \\
O'Neill's, token, 336 & 52{,}864 & 0.22 $\pm$ 0.01 & 1.24 $\pm$ 0.03 & 0.22 $\pm$ 0.01 & 0.4\% $\to$ 71\% \\
O'Neill's, latent, 192 & 5{,}421 & 0.47 $\pm$ 0.005 & 0.78 $\pm$ 0.01 & 0.47 $\pm$ 0.01 & n/a \\
O'Neill's, latent, 336 & 8{,}942 & 0.33 $\pm$ 0.01 & 0.71 $\pm$ 0.004 & 0.33 $\pm$ 0.01 & n/a \\
\bottomrule
\end{tabular}
}
\end{table}

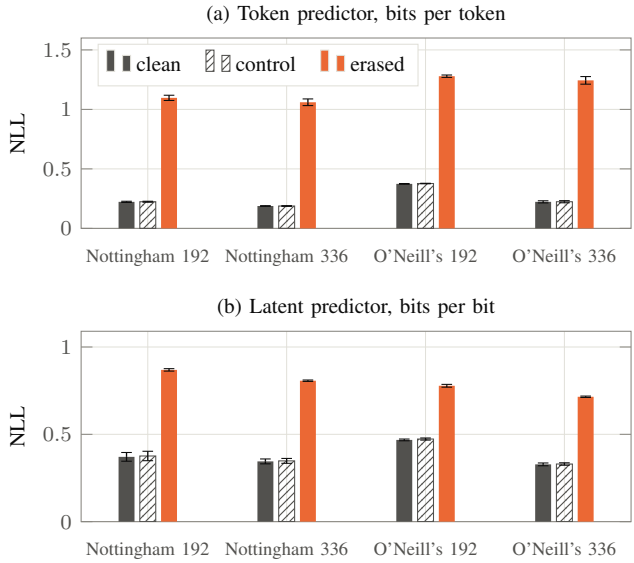
\begin{figure}[t]
\centering
% generated by scripts/make_paper_tables.py; do not edit
\definecolor{clat}{HTML}{2A78D6}
\definecolor{ctok}{HTML}{EB6834}
\definecolor{cink}{HTML}{0B0B0B}
\definecolor{cink2}{HTML}{52514E}
\definecolor{cmuted}{HTML}{898781}
\definecolor{cgrid}{HTML}{E1E0D9}
\begin{tikzpicture}
\begin{axis}[name=e0,  width=\columnwidth, height=4.1cm, ybar, bar width=6pt, ymin=0, ymax=1.60,
  xtick={0,1,2,3}, xticklabels={Nottingham 192,Nottingham 336,O'Neill's 192,O'Neill's 336},
  enlarge x limits=0.16, xtick style={draw=none}, x tick label style={font=\scriptsize},
  ylabel={NLL}, title={(a) Token predictor, bits per token}, title style={font=\footnotesize, text=cink, yshift=-1ex}, legend pos=north west, legend columns=3, legend style={font=\footnotesize, draw=cgrid, fill=white, text=cink, /tikz/every even column/.append style={column sep=0.25cm}},
  grid=major, grid style={cgrid, line width=0.3pt}, axis line style={cmuted}, tick style={cmuted}, tick label style={font=\footnotesize, text=cink2}, label style={font=\footnotesize, text=cink}, error bars/y dir=both, error bars/y explicit, error bars/error bar style={cink, line width=0.5pt}]
\addplot[fill=cink2, draw=none] coordinates {(0,0.223) +- (0,0.005) (1,0.188) +- (0,0.003) (2,0.374) +- (0,0.003) (3,0.223) +- (0,0.009)};
\addlegendentry{clean}
\addplot[pattern=north east lines, pattern color=cink2, draw=cink2, line width=0.4pt] coordinates {(0,0.224) +- (0,0.004) (1,0.188) +- (0,0.003) (2,0.377) +- (0,0.002) (3,0.224) +- (0,0.009)};
\addlegendentry{control}
\addplot[fill=ctok, draw=none] coordinates {(0,1.097) +- (0,0.022) (1,1.060) +- (0,0.028) (2,1.279) +- (0,0.010) (3,1.244) +- (0,0.032)};
\addlegendentry{erased}
\end{axis}
\begin{axis}[name=e1, at={(e0.below south west)}, anchor=north west, yshift=-0.75cm, width=\columnwidth, height=4.1cm, ybar, bar width=6pt, ymin=0, ymax=1.09,
  xtick={0,1,2,3}, xticklabels={Nottingham 192,Nottingham 336,O'Neill's 192,O'Neill's 336},
  enlarge x limits=0.16, xtick style={draw=none}, x tick label style={font=\scriptsize},
  ylabel={NLL}, title={(b) Latent predictor, bits per bit}, title style={font=\footnotesize, text=cink, yshift=-1ex}, 
  grid=major, grid style={cgrid, line width=0.3pt}, axis line style={cmuted}, tick style={cmuted}, tick label style={font=\footnotesize, text=cink2}, label style={font=\footnotesize, text=cink}, error bars/y dir=both, error bars/y explicit, error bars/error bar style={cink, line width=0.5pt}]
\addplot[fill=cink2, draw=none] coordinates {(0,0.371) +- (0,0.025) (1,0.345) +- (0,0.014) (2,0.468) +- (0,0.005) (3,0.328) +- (0,0.008)};

\addplot[pattern=north east lines, pattern color=cink2, draw=cink2, line width=0.4pt] coordinates {(0,0.376) +- (0,0.027) (1,0.348) +- (0,0.014) (2,0.473) +- (0,0.006) (3,0.330) +- (0,0.007)};

\addplot[fill=ctok, draw=none] coordinates {(0,0.869) +- (0,0.007) (1,0.807) +- (0,0.004) (2,0.778) +- (0,0.008) (3,0.715) +- (0,0.004)};

\end{axis}
\end{tikzpicture}
\caption{The erase test of Table~\ref{tab:copy} as bars: test NLL of the targets with a repeat in view when the context is left as it is (clean), when as many unrelated spans are overwritten (control), and when every earlier occurrence of the target's 16-token history is overwritten (erased), for (a) the token predictor and (b) the latent predictor at 192 and 336 tokens. Error bars are the seed SD.}
\label{fig:erase}
\end{figure}

The preceding association between repeat availability and NLL reduction is correlational. To test whether the predictors causally depend on the earlier occurrence, we intervene on the context. For each Nottingham target with an in-view repeat whose earlier occurrence does not overlap the 16 tokens immediately preceding the target, we overwrite every earlier occurrence of that 16-token history together with the token that followed it, using the corresponding positions from a window drawn from another tune. The 16 tokens immediately preceding the target are left unchanged. As a matched control, we overwrite the same number of equally long spans at random context positions that overlap neither an erased occurrence, another control span, nor the target's immediate 16-token history; targets lacking sufficient non-overlapping positions are excluded. For the latent predictor, the edited span additionally includes the six tokens following the occurrence, after which the modified window is encoded normally.

Table~\ref{tab:copy} and Fig.~\ref{fig:erase} summarize the intervention. For the 336-token token predictor, targeted erasure increases NLL on repeated-history targets from \R{cor/t/336/clean} to \R{cor/t/336/erased} bits per token, approaching the \R{nott/lag/t/336/none} bits observed on targets without repeats, whereas the matched control remains at \R{cor/t/336/control}. For the 336-token latent predictor, erasure increases NLL from \R{cor/l/336/clean} to \R{cor/l/336/erased} bits per bit, while the control remains at \R{cor/l/336/control}. A complementary intervention modifies only the token following each earlier occurrence, replacing it with a pitch sampled uniformly from pitches present in the context other than rest, the true target, or an existing continuation token. The model's mean probability assigned to the injected pitch rises from \R{cor/t/336/pnew/clean}\% to \R{cor/t/336/pnew/swap}\%, demonstrating that the predictor uses the earlier occurrence to determine its continuation. Because donor windows, control locations, and replacement pitches are stochastic, we repeat the intervention across \R{cor/draws/n} independent draws. For Nottingham, the erased NLL of the 336-token token predictor ranges only from \R{cor/draws/t/erased/min} to \R{cor/draws/t/erased/max} bits, the control from \R{cor/draws/t/control/min} to \R{cor/draws/t/control/max}, and the swapped-pitch probability from \R{cor/draws/t/pswap/min}\% to \R{cor/draws/t/pswap/max}\%. The corresponding O'Neill's ranges are \R{cor/oneill/draws/t/erased/min} to \R{cor/oneill/draws/t/erased/max} bits, \R{cor/oneill/draws/t/control/min} to \R{cor/oneill/draws/t/control/max} bits, and \R{cor/oneill/draws/t/pswap/min}\% to \R{cor/oneill/draws/t/pswap/max}\%, showing that the conclusion is insensitive to the random draw.

To quantify how much of the long-context improvement can be explained by explicit retrieval, we construct a simple copy baseline. Analogous to cache and nearest-neighbor language models that interpolate a parametric predictor with a distribution retrieved from earlier context~\cite{grave2017cache,khandelwal2020knnlm}, the baseline mixes the 6-token predictor with a point mass on the token following the nearest earlier history match. It backs off from 32-token to 16-token to 8-token matches, and the interpolation weight for each match length is fitted on validation windows. The baseline uses no more history than the 336-token Transformer. It achieves \R{nott/copybase/backoff} bits per token, compared with \R{nott/copybase/base} for the 6-token predictor and \R{nott/tok/336} for the 336-token Transformer, thereby recovering \R{nott/copybase/backoff/gain}\% of the Transformer's NLL reduction. Restricting the baseline to 16-token matches recovers \R{nott/copybase/16/gain}\%. On MAESTRO, the same baseline recovers only \R{maes/copybase/backoff/gain}\%.

\subsection{One trained model with less history}
\label{sec:ablate}

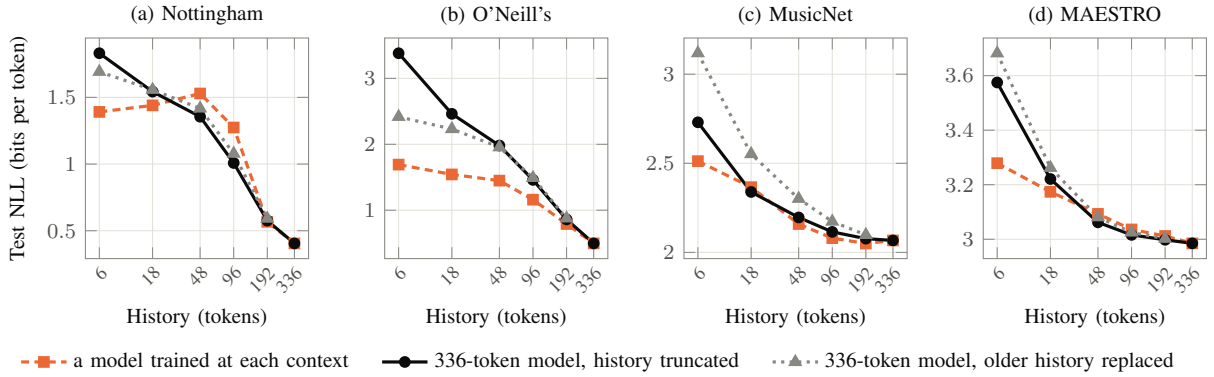
\begin{figure*}[t]
\centering
% generated by scripts/make_paper_tables.py; do not edit
\definecolor{clat}{HTML}{2A78D6}
\definecolor{ctok}{HTML}{EB6834}
\definecolor{cink}{HTML}{0B0B0B}
\definecolor{cink2}{HTML}{52514E}
\definecolor{cmuted}{HTML}{898781}
\definecolor{cgrid}{HTML}{E1E0D9}
\begin{tikzpicture}
\begin{axis}[name=a0,  width=0.25\textwidth, height=4.5cm, xmode=log, log basis x=2,
  xmin=4.5, xmax=450, ymin=0.291, ymax=1.945, xtick={6,18,48,96,192,336}, xticklabels={6,18,48,96,192,336},
  x tick label style={font=\scriptsize, rotate=45, anchor=north east, inner sep=1.5pt},
  ylabel={Test NLL (bits per token)}, xlabel={History (tokens)},
  title={(a) Nottingham}, title style={font=\footnotesize, text=cink, yshift=-1ex}, legend to name=abllegend, legend columns=-1, legend style={font=\footnotesize, draw=none, text=cink, /tikz/every even column/.append style={column sep=0.35cm}},
  grid=major, grid style={cgrid, line width=0.3pt}, axis line style={cmuted}, tick style={cmuted}, tick label style={font=\footnotesize, text=cink2}, label style={font=\footnotesize, text=cink}]
\addplot[ctok, line width=1.2pt, densely dashed, mark=square*, mark size=1.6pt, mark options={solid}] coordinates {(6,1.3910) (18,1.4400) (48,1.5290) (96,1.2740) (192,0.5670) (336,0.4050)};
\addlegendentry{a model trained at each context}
\addplot[cink, line width=1.2pt, mark=*, mark size=1.6pt] coordinates {(6,1.8310) (18,1.5430) (48,1.3550) (96,1.0080) (192,0.5760) (336,0.4050)};
\addlegendentry{336-token model, history truncated}
\addplot[cmuted, line width=1.2pt, dotted, mark=triangle*, mark size=1.9pt, mark options={solid}] coordinates {(6,1.6920) (18,1.5570) (48,1.4170) (96,1.0780) (192,0.5910)};
\addlegendentry{336-token model, older history replaced}
\end{axis}
\begin{axis}[name=a1, at={(a0.right of south east)}, anchor=south west, xshift=1.0cm, width=0.25\textwidth, height=4.5cm, xmode=log, log basis x=2,
  xmin=4.5, xmax=450, ymin=0.266, ymax=3.612, xtick={6,18,48,96,192,336}, xticklabels={6,18,48,96,192,336},
  x tick label style={font=\scriptsize, rotate=45, anchor=north east, inner sep=1.5pt},
   xlabel={History (tokens)},
  title={(b) O'Neill's}, title style={font=\footnotesize, text=cink, yshift=-1ex}, 
  grid=major, grid style={cgrid, line width=0.3pt}, axis line style={cmuted}, tick style={cmuted}, tick label style={font=\footnotesize, text=cink2}, label style={font=\footnotesize, text=cink}]
\addplot[ctok, line width=1.2pt, densely dashed, mark=square*, mark size=1.6pt, mark options={solid}] coordinates {(6,1.6920) (18,1.5430) (48,1.4490) (96,1.1600) (192,0.7910) (336,0.4970)};

\addplot[cink, line width=1.2pt, mark=*, mark size=1.6pt] coordinates {(6,3.3810) (18,2.4620) (48,1.9810) (96,1.4600) (192,0.8610) (336,0.4970)};

\addplot[cmuted, line width=1.2pt, dotted, mark=triangle*, mark size=1.9pt, mark options={solid}] coordinates {(6,2.4170) (18,2.2320) (48,1.9560) (96,1.4910) (192,0.8770)};

\end{axis}
\begin{axis}[name=a2, at={(a1.right of south east)}, anchor=south west, xshift=1.0cm, width=0.25\textwidth, height=4.5cm, xmode=log, log basis x=2,
  xmin=4.5, xmax=450, ymin=1.965, ymax=3.203, xtick={6,18,48,96,192,336}, xticklabels={6,18,48,96,192,336},
  x tick label style={font=\scriptsize, rotate=45, anchor=north east, inner sep=1.5pt},
   xlabel={History (tokens)},
  title={(c) MusicNet}, title style={font=\footnotesize, text=cink, yshift=-1ex}, 
  grid=major, grid style={cgrid, line width=0.3pt}, axis line style={cmuted}, tick style={cmuted}, tick label style={font=\footnotesize, text=cink2}, label style={font=\footnotesize, text=cink}]
\addplot[ctok, line width=1.2pt, densely dashed, mark=square*, mark size=1.6pt, mark options={solid}] coordinates {(6,2.5120) (18,2.3660) (48,2.1590) (96,2.0800) (192,2.0500) (336,2.0670)};

\addplot[cink, line width=1.2pt, mark=*, mark size=1.6pt] coordinates {(6,2.7300) (18,2.3390) (48,2.1960) (96,2.1150) (192,2.0770) (336,2.0670)};

\addplot[cmuted, line width=1.2pt, dotted, mark=triangle*, mark size=1.9pt, mark options={solid}] coordinates {(6,3.1180) (18,2.5520) (48,2.3010) (96,2.1730) (192,2.0980)};

\end{axis}
\begin{axis}[name=a3, at={(a2.right of south east)}, anchor=south west, xshift=1.0cm, width=0.25\textwidth, height=4.5cm, xmode=log, log basis x=2,
  xmin=4.5, xmax=450, ymin=2.929, ymax=3.738, xtick={6,18,48,96,192,336}, xticklabels={6,18,48,96,192,336},
  x tick label style={font=\scriptsize, rotate=45, anchor=north east, inner sep=1.5pt},
   xlabel={History (tokens)},
  title={(d) MAESTRO}, title style={font=\footnotesize, text=cink, yshift=-1ex}, 
  grid=major, grid style={cgrid, line width=0.3pt}, axis line style={cmuted}, tick style={cmuted}, tick label style={font=\footnotesize, text=cink2}, label style={font=\footnotesize, text=cink}]
\addplot[ctok, line width=1.2pt, densely dashed, mark=square*, mark size=1.6pt, mark options={solid}] coordinates {(6,3.2790) (18,3.1740) (48,3.0930) (96,3.0360) (192,3.0120) (336,2.9850)};

\addplot[cink, line width=1.2pt, mark=*, mark size=1.6pt] coordinates {(6,3.5750) (18,3.2210) (48,3.0620) (96,3.0160) (192,2.9980) (336,2.9850)};

\addplot[cmuted, line width=1.2pt, dotted, mark=triangle*, mark size=1.9pt, mark options={solid}] coordinates {(6,3.6820) (18,3.2620) (48,3.0810) (96,3.0250) (192,3.0020)};

\end{axis}
\end{tikzpicture}
\par\vspace{2pt}\ref{abllegend}
\caption{One trained model with less history. Test NLL of the 336-token token predictors when their history is limited at test time to the most recent tokens, by giving the model only those tokens (truncated) or by replacing every older token with a window of another piece (replaced), beside the models trained separately at each context length. Mean over three seeds, on the targets of the sweep (every fourth target for MAESTRO); each panel has its own scale.}
\label{fig:ablate}
\end{figure*}

The sweep compares independently trained models at different maximum context lengths. To measure the dependence of a single trained model on distant history, we take the 336-token predictors and restrict their test-time history to the most recent $R$ tokens for the same six values of $R$. We use two interventions. In the \emph{truncated} condition, the model receives only the final $R$ tokens. In the \emph{replaced} condition, the input retains its full 336-token length, but every position older than $R$ is replaced by the token at the corresponding position of a window from another piece. No model is retrained for these interventions. Because the models were never trained on shortened or mixed histories, performance at small $R$ reflects both the loss of distant information and robustness to an out-of-distribution input format.

Fig.~\ref{fig:ablate} reports the token-predictor results. On Nottingham, the 336-token model achieves \R{nott/abl/t/trunc/336} bits per token with its full history. Restricting history to 192 tokens raises NLL to \R{nott/abl/t/trunc/192} under truncation and \R{nott/abl/t/repl/192} under replacement, close to the \R{nott/tok/192} bits attained by models trained directly with 192-token context. With only 96 recent tokens, NLL rises further to \R{nott/abl/t/trunc/96} and \R{nott/abl/t/repl/96}. The degradation is concentrated on targets whose repeat is visible at 336 tokens: their NLL increases from \R{nott/abl/t/trunc/336/repeat} bits with full history to \R{nott/abl/t/trunc/192/repeat} with 192 tokens and \R{nott/abl/t/trunc/96/repeat} with 96, whereas targets without a repeat increase from \R{nott/abl/t/trunc/336/none} to \R{nott/abl/t/trunc/192/none} and \R{nott/abl/t/trunc/96/none}. Thus a single trained model loses most of its long-context advantage when distant history is removed by either intervention. At 6 and 18 tokens, the replaced-history model performs worse than models trained directly at those context lengths (\R{nott/abl/t/repl/6} and \R{nott/abl/t/repl/18} bits versus \R{nott/tok/6} and \R{nott/tok/18}), consistent with the distribution shift introduced by severe history restriction. At 96 tokens, it outperforms some separately trained models because not all of those runs discovered the copying solution, and at 48 tokens it outperforms the separately trained models, which score worse than the 6-token predictor (Section~\ref{sec:stability}).

O'Neill's exhibits the same dependence on distant history. The full-history model scores \R{oneill/abl/t/trunc/336} bits; with only the most recent 192 tokens, NLL becomes \R{oneill/abl/t/trunc/192} under truncation and \R{oneill/abl/t/repl/192} under replacement, compared with \R{oneill/tok/192} for models trained directly at 192 tokens. With 96 recent tokens, the values are \R{oneill/abl/t/trunc/96} and \R{oneill/abl/t/repl/96}, compared with \R{oneill/tok/96} for the 96-token sweep models. On repeated-history targets, NLL increases from \R{oneill/abl/t/trunc/336/repeat} to \R{oneill/abl/t/trunc/192/repeat} and \R{oneill/abl/t/trunc/96/repeat}. The restricted O'Neill's model performs worse than separately trained models at every shorter context length; unlike Nottingham, the O'Neill's 96-token sweep models are also stable across seeds (Section~\ref{sec:oneills}). In contrast, restricting the 336-token MusicNet and MAESTRO models to 96 tokens has only a small effect: MusicNet changes from \R{music/abl/t/trunc/336} to \R{music/abl/t/trunc/96} bits and MAESTRO from \R{maes/abl/t/trunc/336} to \R{maes/abl/t/trunc/96}, close to the corresponding 96-token sweep values (\R{music/abl/t/sweep/96} and \R{maes/abl/t/sweep/96}). This is consistent with the independently trained sweep, in which most of the benefit on these two corpora is already attained by 96 tokens. MAESTRO is evaluated on every fourth target in this analysis.

The latent predictor shows the same qualitative dependence when older latent steps are replaced. On Nottingham, NLL is \R{nott/abl/l/trunc/336} bits per bit with full history, \R{nott/abl/l/repl/192} with the most recent 192 tokens preserved, and \R{nott/abl/l/repl/96} with 96 tokens preserved; the independently trained models at 192 and 96 tokens score \R{nott/lat/192} and \R{nott/lat/96}, respectively. Direct truncation produces NLL above 1 bit per bit at 6 and 18 tokens, so we report only the replacement intervention for the latent predictor.

\subsection{A second folk corpus}
\label{sec:oneills}

We repeat the protocol on the O'Neill's test tunes long enough to provide a complete window (Section~\ref{sec:data}), comprising \R{oneill/long/test} of \R{oneill/pieces/test} test tunes. The token predictor decreases from \R{oneill/tok/6} bits per token at 6 tokens to \R{oneill/tok/96} at 96, \R{oneill/tok/192} at 192, and \R{oneill/tok/336} at 336, a \R{oneill/tok/drop}\% reduction (Nottingham: \R{nott/tok/drop}\%). The absolute decrease is \R{oneill/boot/t/6-336} bits, with a 95\% interval of \R{oneill/boot/t/6-336/lo} to \R{oneill/boot/t/6-336/hi}. The latent predictor decreases from \R{oneill/lat/6} to \R{oneill/lat/336} bits per bit, a \R{oneill/lat/drop}\% reduction (Nottingham: \R{nott/lat/drop}\%). Two differences from Nottingham are notable: token NLL does not increase from 6 to 48 tokens, and the three seeds agree at 96 tokens (SD \R{oneill/tok/96/sd} bits versus \R{nott/tok/96/sd} on Nottingham), so we do not observe the same seed-dependent split between runs that do and do not discover copying. We did not perform the fixed-budget experiment of Section~\ref{sec:stability} on O'Neill's.

The repeat analysis reproduces the Nottingham pattern. At 336 tokens, \R{oneill/inview/336}\% of targets have an in-view repeat (Nottingham: \R{nott/inview/336}\%), and the token predictor has attained \R{oneill/reached/t/96}\%, \R{oneill/reached/t/192}\%, and 100\% of its total reduction at 96, 192, and 336 tokens. Repeated-history targets account for \R{oneill/redshare/t/repeat}\% of the token-predictor reduction and \R{oneill/redshare/l/repeat}\% of the latent-predictor reduction, and for \R{oneill/copy/t/all}\% of these targets the continuation of the earlier occurrence exactly matches the target. The copy baseline recovers \R{oneill/copybase/backoff/gain}\% of the token predictor's reduction (Nottingham: \R{nott/copybase/backoff/gain}\%). In the erase intervention (Table~\ref{tab:copy}), removing the earlier occurrences increases the 336-token token predictor's NLL on these targets from \R{cor/oneill/t/336/clean} to \R{cor/oneill/t/336/erased} bits per token, whereas the matched control remains at \R{cor/oneill/t/336/control}. Replacing only the continuation token increases the probability assigned to the injected pitch from \R{cor/oneill/t/336/pnew/clean}\% to \R{cor/oneill/t/336/pnew/swap}\%. For the 336-token latent predictor, erasure raises NLL from \R{cor/oneill/l/336/clean} to \R{cor/oneill/l/336/erased} bits per bit, while the control remains at \R{cor/oneill/l/336/control}.

\subsection{A corpus in between}
\label{sec:musicnet}

MusicNet contains substantially fewer exact repeats than the two folk corpora but more than MAESTRO: \R{music/inview/336}\% of its targets have a 16-token repeat in view at 336 tokens, compared with \R{maes/inview/336}\% for MAESTRO and \R{nott/inview/336}\% and \R{oneill/inview/336}\% for Nottingham and O'Neill's. Its token NLL decreases from \R{music/tok/6} bits per token at 6 tokens to \R{music/tok/96} at 96 and \R{music/tok/336} at 336, a \R{music/tok/drop}\% reduction; the absolute decrease is \R{music/boot/t/6-336} bits (95\% interval \R{music/boot/t/6-336/lo} to \R{music/boot/t/6-336/hi}). By 96 tokens, the model has attained \R{music/reached/t/96}\% of this reduction. The 336-token models are slightly worse than the 192-token models by \R{music/boot/t/336-192} bits (interval \R{music/boot/t/336-192/lo} to \R{music/boot/t/336-192/hi}), providing no evidence of additional benefit from the longest tested context. Targets with an in-view repeat account for \R{music/redshare/t/repeat}\% of the total reduction and targets without one for \R{music/redshare/t/none}\%. The copy baseline recovers \R{music/copybase/backoff/gain}\% of the reduction, intermediate between MAESTRO's \R{maes/copybase/backoff/gain}\% and the folk corpora's \R{nott/copybase/backoff/gain}\% and \R{oneill/copybase/backoff/gain}\%. This exceeds the contribution attributed to 16-token repeats because the baseline also retrieves 8-token matches: a 32-, 16-, or 8-token match exists for \R{music/copybase/backoff/matched}\% of targets, on which the baseline averages \R{music/copybase/m/bits_copy} bits compared with \R{music/copybase/m/bits_T336} for the 336-token Transformer and \R{music/copybase/m/bits_T6} for the 6-token model. On these matched targets the Transformer obtains \R{music/copybase/m/t336}\% of its total reduction, whereas the baseline's gain on them equals \R{music/copybase/m/copy}\% of that total. Thus the baseline quantifies the potential benefit of exact retrieval on MusicNet, but does not establish that the Transformer realizes its improvement through the same mechanism. On Nottingham and O'Neill's, the corresponding quantities are closely aligned: \R{nott/copybase/m/t336}\% for the Transformer versus \R{nott/copybase/m/copy}\% for the baseline on Nottingham, and \R{oneill/copybase/m/t336}\% versus \R{oneill/copybase/m/copy}\% on O'Neill's. The MusicNet sweep uses three seeds, token predictors only, and \R{music/pieces/testwin} test pieces; consequently, its bootstrap intervals are comparatively wide, and the 6-token seed SD is \R{music/tok/6/sd} bits.

\subsection{Copying depends on training}
\label{sec:stability}

\begin{table*}[t]
\centering
\caption{Nottingham under three setups, test NLL at four context lengths (tokens), mean and SD over three seeds. Latent values in bits per bit, token values in bits per token. The setups differ in tokenizer, window, autoencoder and transposition at once, so values are comparable within a row.}
\label{tab:variants}
% generated by scripts/make_paper_tables.py; do not edit
\begin{tabular}{lcccc}
\toprule
\textbf{Setup} & \textbf{6} & \textbf{96} & \textbf{192} & \textbf{336} \\
\midrule
Legacy grid, 480-token windows, $D{=}32$, latent & 0.80 $\pm$ 0.0002 & 0.80 $\pm$ 0.0005 & 0.78 $\pm$ 0.001 & 0.72 $\pm$ 0.01 \\
Legacy grid, 480-token windows, $D{=}32$, token & 1.22 $\pm$ 0.001 & 1.20 $\pm$ 0.004 & 1.04 $\pm$ 0.01 & 0.84 $\pm$ 0.02 \\
Score grid, 384-token windows, $D{=}64$, latent & 0.91 $\pm$ 0.001 & 0.91 $\pm$ 0.003 & 0.77 $\pm$ 0.004 & 0.63 $\pm$ 0.01 \\
Score grid, 384-token windows, $D{=}64$, token & 1.39 $\pm$ 0.004 & 1.09 $\pm$ 0.02 & 1.15 $\pm$ 0.02 & 0.81 $\pm$ 0.25 \\
Score grid, as above, with transposition, latent & 0.86 $\pm$ 0.002 & 0.78 $\pm$ 0.004 & 0.57 $\pm$ 0.01 & 0.47 $\pm$ 0.01 \\
Score grid, as above, with transposition, token & 1.39 $\pm$ 0.01 & 1.27 $\pm$ 0.33 & 0.57 $\pm$ 0.01 & 0.41 $\pm$ 0.01 \\
\bottomrule
\end{tabular}

\end{table*}

\begin{figure}[t]
\centering
\input{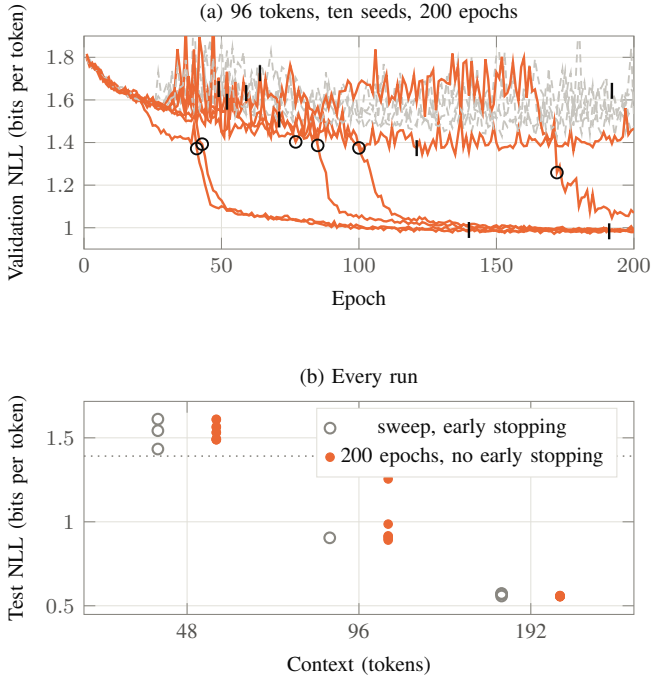}
\caption{Seed spread of the Nottingham token predictor. (a) Validation NLL against epoch for ten runs at 96 tokens trained for 200 epochs with no early stopping; coloured lines drop (Section~\ref{sec:stability}), with a circle at the epoch of the drop, grey dashed lines do not, and a tick marks where the sweep's early stopping would have ended each run. (b) Test NLL of every run: the three early-stopped runs of the sweep (circles) and the ten fixed-budget runs (dots). The dotted line is the 6-token predictor.}
\label{fig:curves}
\end{figure}

Within the main sweep, the Nottingham token predictor at 96 tokens has mean test NLL \R{nott/tok/96} bits per token with seed SD \R{nott/tok/96/sd}. One seed reaches \R{sweepseed/96/s0/test} bits and scores \R{sweepseed/96/s0/inview} bits on targets with an in-view repeat versus \R{sweepseed/96/s0/none} on targets without one. The other two seeds exhibit much smaller repeat-specific gains, scoring \R{sweepseed/96/s7/inview} and \R{sweepseed/96/s42/inview} bits on repeated-history targets versus \R{sweepseed/96/s7/none} and \R{sweepseed/96/s42/none} on targets without repeats; early stopping terminates these runs at epochs \R{sweepseed/96/s7/stop} and \R{sweepseed/96/s42/stop}.

To disentangle seed effects from the stopping criterion, we train ten additional token predictors at each of 48, 96, and 192 tokens (seeds 0 to 9) for a fixed budget of 200 epochs without early stopping, retaining the checkpoint with the lowest validation NLL (Fig.~\ref{fig:curves}). We define a \emph{drop} as a decrease in validation NLL of at least 0.2 bits below the minimum attained during the first 20 epochs. At 96 tokens, \R{seeds/96/drop/n} of ten runs undergo such a transition after an extended plateau, at epochs \R{seeds/96/drop/min} to \R{seeds/96/drop/max}. Five finish between \R{seeds/96/drop/nll/min} and \R{seeds/96/drop/nll/second} bits per token and one at \R{seeds/96/drop/nll/max}; the remaining four finish between \R{seeds/96/nodrop/nll/min} and \R{seeds/96/nodrop/nll/max}. Those four still perform better on repeated-history targets (\R{seeds/96/nodrop/inview/min} to \R{seeds/96/nodrop/inview/max} bits) than on targets without repeats (\R{seeds/96/nodrop/none/min} to \R{seeds/96/nodrop/none/max}). Applied retrospectively to these learning curves, the main sweep's early-stopping rule would terminate \R{seeds/96/escut} of the \R{seeds/96/drop/n} eventually dropping runs before their transition, and would stop the four runs that never drop between epochs \R{seeds/96/nodrop/es/min} and \R{seeds/96/nodrop/es/max}. All ten runs achieve their best validation NLL between epochs \R{seeds/96/best/min} and \R{seeds/96/best/max}; because several are still improving at 200 epochs, additional training could change the fraction that reaches the lower-NLL solution. At 48 tokens, where only \R{nott/inview/48}\% of targets have an in-view repeat, all ten runs score \R{seeds/48/inview/min} to \R{seeds/48/inview/max} bits on repeated-history targets and finish between \R{seeds/48/min} and \R{seeds/48/max} bits overall, above the \R{nott/tok/6} bits of the 6-token predictor.

At 192 tokens, all \R{seeds/192/drop/n} fixed-budget runs exhibit the transition, between epochs \R{seeds/192/drop/min} and \R{seeds/192/drop/max}, and finish between \R{seeds/192/min} and \R{seeds/192/max} bits per token. Likewise, every 192- and 336-token run in the main sweep drops between epochs \R{sweepdrop/long/min} and \R{sweepdrop/long/max}. Thus, at 96 tokens, acquisition of the copying solution depends strongly on both initialization and training duration, with only about half of the runs reaching it within 200 epochs; at 192 tokens, all tested runs reach the lower-NLL solution and converge closely across seeds. The abrupt transition is qualitatively reminiscent of the emergence of induction heads in language models~\cite{olsson2022induction}. Accordingly, the three-seed means at 48 and 96 tokens should be interpreted in light of this bimodal optimization behavior.

Table~\ref{tab:variants} assesses sensitivity to the broader experimental setup. Without transposition, every Nottingham predictor reaches its best validation NLL within \R{nottnoaug/lat/ep/max} epochs, and the 336-token token predictor has mean NLL \R{nottnoaug/tok/336} bits with SD \R{nottnoaug/tok/336/sd}, reflecting one seed that learns substantially stronger copying than the other two. With transposition, the 192- and 336-token models reach their best validation NLL after \R{nott/longep/min} to \R{nott/longep/max} epochs, every seed acquires copying at these context lengths, and the 336-token SD falls to \R{nott/tok/336/sd}. The latent predictor is comparatively stable across all setups, with no seed SD above \R{nott/lat/maxsd}. Despite differences in absolute values, all three setups preserve the same qualitative context-length pattern: small changes at short context and a substantially larger reduction by 336 tokens, including the first rerun using the legacy tokenizer and 480-token windows.

\subsection{An averaging predictor does not use long context}
\label{sec:v5}

\begin{table}[t]
\centering
\caption{The causal latent predictor against v5, the averaging predictor of our earlier draft, on the same latents. Test NLL in bits per bit, mean over three seeds.}
\label{tab:v5}
\resizebox{\columnwidth}{!}{% generated by scripts/make_paper_tables.py; do not edit
\begin{tabular}{rcccc}
\toprule
 & \multicolumn{2}{c}{\textbf{Nottingham}} & \multicolumn{2}{c}{\textbf{MAESTRO}} \\
\cmidrule(lr){2-3}\cmidrule(lr){4-5}
\textbf{Context} & Causal & v5 & Causal & v5 \\
\midrule
6 & 0.857 $\pm$ 0.002 & 0.856 $\pm$ 0.0001 & 0.887 $\pm$ 0.0004 & 0.887 $\pm$ 0.0004 \\
18 & 0.842 $\pm$ 0.002 & 0.852 $\pm$ 0.001 & 0.881 $\pm$ 0.00003 & 0.881 $\pm$ 0.0002 \\
48 & 0.834 $\pm$ 0.005 & 0.843 $\pm$ 0.004 & 0.879 $\pm$ 0.0004 & 0.879 $\pm$ 0.0004 \\
96 & 0.775 $\pm$ 0.004 & 0.891 $\pm$ 0.001 & 0.877 $\pm$ 0.0004 & 0.879 $\pm$ 0.0003 \\
192 & 0.571 $\pm$ 0.015 & 0.886 $\pm$ 0.001 & 0.877 $\pm$ 0.00010 & 0.878 $\pm$ 0.0003 \\
336 & 0.465 $\pm$ 0.009 & 0.891 $\pm$ 0.001 & 0.877 $\pm$ 0.0004 & 0.878 $\pm$ 0.0003 \\
\bottomrule
\end{tabular}
}
\end{table}

Table~\ref{tab:v5} compares v5 with the causal latent predictor on the same latent representation. Up to 48 tokens, their NLLs differ by about 0.01 bits or less. On Nottingham, v5 subsequently degrades to \R{nott/v5/96} bits per bit at 96 tokens and \R{nott/v5/336} at 336, whereas the causal predictor improves to \R{nott/lat/336}. On MAESTRO, where little additional reduction remains beyond 96 tokens, the two predictors remain close. These results indicate that the mean-pooled architecture fails to exploit long-range repetition, plausibly because averaging over context positions makes retrieval of a specific distant continuation difficult. Apparent context saturation can therefore arise from limitations of the predictor architecture rather than from an absence of useful information in the data.

\subsection{Autoencoder seeds}
\label{sec:aeseeds}

The main latent-context sweeps use one trained autoencoder per corpus. To quantify sensitivity to representation learning, we train two additional Nottingham autoencoders with identical settings (seeds 1 and 2; test-token accuracies \R{aeseed/1/acc}\% and \R{aeseed/2/acc}\%) and, for each autoencoder, one causal latent predictor per context using predictor seed 42, matching the reference run. All three autoencoders yield the same qualitative curve: NLL changes little through 48 tokens and decreases sharply after 96, with 6-to-336-token reductions of \R{aeseed/42/drop}\%, \R{aeseed/1/drop}\%, and \R{aeseed/2/drop}\%. Across autoencoders, the SD ranges from \R{aeseed/sd/min} to \R{aeseed/sd/max} bits per bit, compared with \R{aeseed/psd/min} to \R{aeseed/psd/max} across predictor seeds, and is larger at \R{aeseed/larger} of the six context lengths, most notably at 336 tokens (\R{aeseed/336/sd} versus \R{aeseed/336/psd}). Representation seed therefore contributes more variability to latent NLL than predictor seed in most settings, without changing the qualitative conclusion. We did not repeat this analysis for MAESTRO.

\subsection{Generation}
\label{sec:gen}

% Generation section of main.tex; numbers come from tables/numbers.tex.
Prediction is measured on real histories. This section is a secondary diagnostic with one question: does the copying seen in prediction also appear as exact repetition inside the streams the models generate? We generate from every Nottingham model. The distance defined below compares three surface statistics with the corpus, and we do not read it as a measure of musical quality.

\textbf{Setup.} Each of 100 test tunes primes one stream with its first 336 tokens, the longest context in the sweep, so every model receives the same prime and uses the last $T$ tokens or $k$ latent steps of it. The model then generates 480 tokens. Latent streams are decoded by argmax, so all randomness comes from the predictor. For each model, the sampling temperature $\tau$ is swept on 100 validation tunes until the best value has a tried neighbour on both sides, and that temperature is used once on the test tunes. Because the tuned temperature differs between models, we also generate from every model at $\tau{=}1$, which keeps the temperature fixed across context lengths.

\textbf{Statistics.} For a stream $x_1,\dots,x_N$ of $N{=}480$ tokens we compute three statistics. The adjacent repeat ratio $r$ and the token entropy $H$ are
\[
r = \frac{1}{N-1}\sum_{i=1}^{N-1} \mathbf{1}[x_{i+1} = x_i], \qquad H = -\sum_{v} p_v \log_2 p_v,
\]
where $\mathbf{1}[\cdot]$ is 1 when its condition holds and 0 otherwise, and $p_v$ is the share of the stream's positions that hold token $v$, rests included. The block variety $u$ is the number of distinct token values in a 48-token block, rests included, averaged over the stream's 10 blocks. Our earlier draft called $u$ note density; it counts distinct pitches and rests, and the representation has no onsets, so it cannot count notes. The distance of a set $S$ of streams to the corpus is
\begin{equation}
d(S) = \frac{1}{3}\sum_{s \in \{r, H, u\}} \frac{\lvert \bar{s}_S - \mu_s \rvert}{\sigma_s},
\end{equation}
where $\bar{s}_S$ is the mean of statistic $s$ over the streams of $S$, and $\mu_s$ and $\sigma_s$ are its mean and SD over the reference, the training split cut into 480-token streams. The structural repetition rate of a stream is the share of positions whose 16-token history, when it is not one held pitch, occurs earlier in the same stream within 336 tokens. It measures the repetition of passages, which $r$ does not: $r$ only counts adjacent tokens that are equal.

\textbf{Real band.} A distance of zero is not the target, because a finite sample of real music does not reproduce the reference exactly. Our earlier draft compared against a single set of 100 corpus sequences, which scored 0.36. Here we draw 2000 sets of 100 test tunes, with replacement, take one stream from each (its first 480 tokens), and report the 5th to 95th percentile of their distance: \R{gen/band/lo} to \R{gen/band/hi}. The sampling unit is the tune, as it is for the generated streams. Real streams have a structural repetition rate of \R{gen/struct/lo} to \R{gen/struct/hi}. Generated streams follow a 336-token prime, so they come from later in a tune than the first 480 tokens, but only \R{gen/longtunes} test tunes have 336 plus 480 tokens, too few to cut real streams at the same place. We therefore also compute the band from the last 480 tokens of each tune (\R{gen/cut/last/lo} to \R{gen/cut/last/hi}) and from one 480-token segment at a random offset in each tune (\R{gen/cut/random/lo} to \R{gen/cut/random/hi}).

\begin{table}[t]
\centering
\caption{Generation from the Nottingham models, 100 test primes per model, mean and SD over the three training seeds. $\tau$ lists the temperatures chosen on validation primes. Repetition is the structural repetition rate. The last column is the distance with $\tau{=}1$ for every model. The first row is the 5th to 95th percentile of 2000 sets of 100 real test tunes, one stream each, cut from the start of each tune.}
\label{tab:gen}
\resizebox{\columnwidth}{!}{% generated by scripts/make_paper_tables.py; do not edit
\begin{tabular}{lcccc}
\toprule
 & \multicolumn{3}{c}{\textbf{Tuned temperature}} & \textbf{$\tau{=}1$} \\
\cmidrule(lr){2-4}\cmidrule(lr){5-5}
\textbf{System} & $\tau$ & \textbf{Distance} & \textbf{Repetition} & \textbf{Distance} \\
\midrule
\emph{Real test tunes, 100 per set (5th to 95th pct.)} & n/a & \emph{0.03 to 0.17} & \emph{0.48 to 0.52} & n/a \\
AE round trip, all test streams & n/a & 0.07 & 0.49 & n/a \\
\midrule
Latent, 6 tokens & 0.2, 0.3 & 1.21 $\pm$ 0.19 & 0.01 $\pm$ 0.02 & 5.97 $\pm$ 0.03 \\
Latent, 18 tokens & 0.3, 0.4 & 1.19 $\pm$ 0.19 & 0.00 $\pm$ 0.002 & 5.51 $\pm$ 0.15 \\
Latent, 48 tokens & 0.3, 0.4 & 1.13 $\pm$ 0.16 & 0.01 $\pm$ 0.002 & 5.33 $\pm$ 0.15 \\
Latent, 96 tokens & 0.4, 0.5 & 1.25 $\pm$ 0.07 & 0.02 $\pm$ 0.003 & 5.20 $\pm$ 0.17 \\
Latent, 192 tokens & 0.3, 0.4, 0.5 & 0.88 $\pm$ 0.09 & 0.12 $\pm$ 0.08 & 4.91 $\pm$ 0.12 \\
Latent, 336 tokens & 0.1, 0.3 & 0.68 $\pm$ 0.08 & 0.35 $\pm$ 0.08 & 4.08 $\pm$ 0.18 \\
\midrule
Token, 6 tokens & 0.7, 0.8 & 1.49 $\pm$ 0.09 & 0.01 $\pm$ 0.004 & 2.83 $\pm$ 0.07 \\
Token, 18 tokens & 0.8 & 1.66 $\pm$ 0.06 & 0.01 $\pm$ 0.002 & 2.33 $\pm$ 0.01 \\
Token, 48 tokens & 0.9 & 1.37 $\pm$ 0.04 & 0.00 $\pm$ 0.0007 & 2.21 $\pm$ 0.05 \\
Token, 96 tokens & 0.8, 0.9, 1 & 1.09 $\pm$ 0.70 & 0.06 $\pm$ 0.09 & 1.78 $\pm$ 0.24 \\
Token, 192 tokens & 0.8 & 0.25 $\pm$ 0.06 & 0.41 $\pm$ 0.01 & 1.92 $\pm$ 0.12 \\
Token, 336 tokens & 0.8, 0.9 & 0.29 $\pm$ 0.01 & 0.41 $\pm$ 0.14 & 1.69 $\pm$ 0.28 \\
\bottomrule
\end{tabular}
}
\end{table}

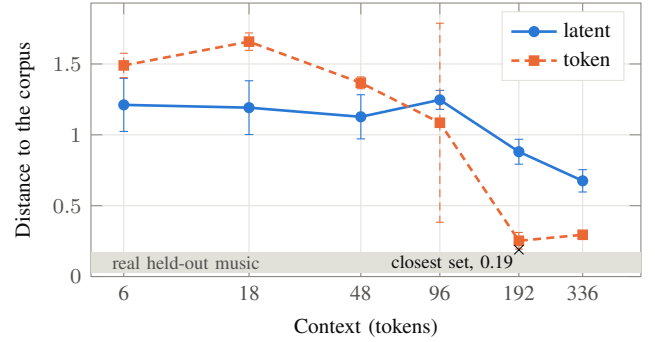
\begin{figure}[t]
\centering
% generated by scripts/make_paper_tables.py; do not edit
\definecolor{clat}{HTML}{2A78D6}
\definecolor{ctok}{HTML}{EB6834}
\definecolor{cink}{HTML}{0B0B0B}
\definecolor{cink2}{HTML}{52514E}
\definecolor{cmuted}{HTML}{898781}
\definecolor{cgrid}{HTML}{E1E0D9}
\begin{tikzpicture}
\begin{axis}[width=\columnwidth, height=5.2cm, xmode=log, log basis x=2,
  xmin=4.5, xmax=560, ymin=0, ymax=1.93, xtick={6,18,48,96,192,336}, xticklabels={6,18,48,96,192,336},
  xlabel={Context (tokens)}, ylabel={Distance to the corpus},
  legend pos=north east, legend style={font=\footnotesize, draw=cgrid, fill=white, text=cink},
  grid=major, grid style={cgrid, line width=0.3pt}, axis line style={cmuted}, tick style={cmuted}, tick label style={font=\footnotesize, text=cink2}, label style={font=\footnotesize, text=cink}]
\fill[cgrid] (axis cs:4.5,0.028) rectangle (axis cs:560,0.171);
\node[font=\scriptsize, text=cink2, anchor=west] at (axis cs:5,0.099) {real held-out music};
\addplot[clat, line width=1pt, mark=*, mark size=1.6pt, error bars/.cd, y dir=both, y explicit] coordinates {(6,1.2114) +- (0,0.1873) (18,1.1918) +- (0,0.1898) (48,1.1272) +- (0,0.1557) (96,1.2469) +- (0,0.0671) (192,0.8807) +- (0,0.0878) (336,0.6754) +- (0,0.0787)};
\addlegendentry{latent}
\addplot[ctok, line width=1pt, densely dashed, mark=square*, mark size=1.5pt, mark options={solid}, error bars/.cd, y dir=both, y explicit] coordinates {(6,1.4898) +- (0,0.0859) (18,1.6577) +- (0,0.0614) (48,1.3669) +- (0,0.0417) (96,1.0854) +- (0,0.7022) (192,0.2525) +- (0,0.0588) (336,0.2948) +- (0,0.0130)};
\addlegendentry{token}
\addplot[only marks, mark=x, mark size=2.5pt, cink, forget plot] coordinates {(192,0.1893)};
\node[font=\scriptsize, text=cink, anchor=north east, inner sep=2pt] at (axis cs:192,0.1893) {closest set, 0.19};
\end{axis}
\end{tikzpicture}
\caption{Distance of generated streams to the training-split reference against context length, at the tuned temperature, mean and SD over three training seeds, 100 test primes per model. The shaded band is the 5th to 95th percentile of 2000 sets of 100 real test tunes cut from the start of each tune, and the cross marks the closest single generated set, at 0.19.}
\label{fig:gen}
\end{figure}

Table~\ref{tab:gen} and Fig.~\ref{fig:gen} give the results. Passing real test streams through the autoencoder gives a distance of \R{gen/ae/dist} and \R{gen/ae/acc}\% token accuracy, inside the real band, so codec distortion on encoded real music is negligible under these statistics.

The closest generated set, from one of the \R{gen/min/ctx}-token \R{gen/min/kind} models, scores \R{gen/min/dist}. Of the \R{gen/nsets} generated sets, \R{gen/cut/first/nin} fall inside the band of streams cut from the start of each tune, \R{gen/cut/random/nin} inside the band of random segments, and \R{gen/cut/last/nin} inside the band of streams cut from the end of each tune, which reaches \R{gen/cut/last/hi}. Whether the best set counts as matching real music therefore depends on where the real streams are cut, and we do not claim a match.

From 6 to 48 tokens of context, both predictors produce streams at a distance above 1 and with almost no repeated passages: the latent streams have a repetition rate of at most \R{gen/latent/short/struct/max}, against about \R{gen/struct/real} for real music. At 96 tokens the same holds for the latent models and for the two token models that did not learn to copy; the one that did generates at a distance of \R{gen/token/96/mindist}. The token streams are too repetitive at the level of adjacent tokens and too spread over pitches: at 6 tokens their repeat ratio is \R{gen/token/6/rep} and their token entropy \R{gen/token/6/ent} bits, against \R{gen/ref/rep} and \R{gen/ref/ent} in the reference.

Long context changes this in the same place it changes prediction. At 192 and 336 tokens the token streams reach distances of \R{gen/token/192/dist} and \R{gen/token/336/dist} with repetition rates of \R{gen/token/192/struct} and \R{gen/token/336/struct}. The latent streams improve more slowly, to \R{gen/latent/192/dist} and \R{gen/latent/336/dist} with repetition rates of \R{gen/latent/192/struct} and \R{gen/latent/336/struct}. The long-context models generate passages that recur within the stream, which the short-context models do not do at their chosen temperatures.

At $\tau{=}1$ every model is further from the reference, and the advantage of longer context remains for the token predictor, though the curve is not monotonic: its distance falls from \R{gen/fixed/token/6/dist} at 6 tokens to \R{gen/fixed/token/96/dist} at 96, rises to \R{gen/fixed/token/192/dist} at 192 and reaches its lowest value, \R{gen/fixed/token/336/dist}, at 336. Only the 192 and 336-token models produce recurring passages, with repetition rates of \R{gen/fixed/token/192/struct} and \R{gen/fixed/token/336/struct}. The latent streams at $\tau{=}1$ are far from the reference at every context (\R{gen/fixed/latent/336/dist} at 336 tokens) and almost never repeat a passage.

The latent codes explain part of this. Of the codes of the encoded test split, \R{gen/codes/real/exact}\% occur exactly among the codes of the encoded training split, and the mean Hamming distance to the nearest training code is \R{gen/codes/real/ham} bits. Of the codes that the latent models generate at their tuned temperature, \R{gen/codes/short/tuned/exact/max}\% occur in training at 6 to 48 tokens of context, with the nearest training code at least \R{gen/codes/short/tuned/ham/min} bits away on average, and \R{gen/codes/336/tuned/exact}\% occur at 336 tokens, \R{gen/codes/336/tuned/ham} bits away. At $\tau{=}1$ almost no generated code occurs in training, and the nearest is at least \R{gen/codes/fixed/ham/min} bits away. The factorized predictor samples each bit separately, so it produces combinations of bits that the encoder never produces, and the decoder must map them to tokens. A low temperature moves every bit toward its most likely value and reduces the problem, which is one reason the latent models choose low temperatures.

The latent models at 336 tokens reach their best distance at temperatures as low as $\tau{=}\R{gen/latent/336/tmin}$, where sampling is close to taking the most likely bit. Their streams therefore show the predictor's most likely continuation more than the spread of continuations it assigns probability to, and their distance should not be read as a measure of variety.

% ============================================================
\section{Failure Modes That Can Produce Spurious Context Saturation}
\label{sec:correction}
% ============================================================

An earlier draft reported that Nottingham latent-prediction accuracy increased by 2.9 points from 5 to 16 tokens of context and then remained approximately constant through 341 tokens. A subsequent audit of the code and data identified four failure modes in the prediction experiments and one in the generation evaluation. All are corrected in the experiments reported above. Because the audit did not remove these factors one at a time, and Table~\ref{tab:variants} changes several components jointly, we do not attribute a specific fraction of the earlier curve to any individual failure mode. Instead, we treat each as a plausible source of bias in context-length measurements that should be checked independently.

\textbf{Leaky splits.} Lee et al.~\cite{lee2022dedup} report that more than 4\% of validation examples in standard language-modeling datasets have a near-duplicate in training, and that Transformer-XL perplexity on those examples is approximately half that on the remainder. Analogous leakage occurred in our earlier music datasets. The earlier Nottingham corpus contained three renderings of each tune (full, melody, and chords) and randomly split the renderings, so \R{old/nottingham/tune}\% of test sequences had another rendering of the same tune in training. At the content level, \R{old/nottingham/verbatim}\% of non-constant 48-token test blocks occurred verbatim in the training split. MAESTRO and MusicNet~\cite{thickstun2017musicnet} were first segmented into 480-token chunks and then randomly split, so \R{old/maestro/performance}\% of MAESTRO test chunks and \R{old/musicnet/recording}\% of MusicNet test chunks came from a performance or recording that also contributed training chunks, although only \R{old/maestro/verbatim}\% and \R{old/musicnet/verbatim}\% of test blocks occurred verbatim in training. We obtained these source-level statistics by reconstructing the earlier datasets from the raw files.

\textbf{Latents that overlapped their targets.} In the earlier encoder, latent step $j$ depended on tokens $4j{-}3$ through $4j{+}18$ of its block. Consequently, a predictor of step $j$ conditioned on steps $j{-}k$ through $j{-}1$ indirectly observed tokens that were also encoded by the target step itself. The first few added context steps therefore revealed progressively more target-overlapping information until the overlap was exhausted, creating an artificial short-context improvement. At the first step of each block, where all available context is strictly in the past, the earlier predictor did not improve with context; on tunes absent from training, its performance was no better than the per-bit marginal.

\textbf{An averaging predictor.} The earlier experiments used v5, whose mean-pooled readout fails to benefit from long context even on the corrected data (Section~\ref{sec:v5}).

\textbf{A per-file tempo estimate.} The earlier tokenizer placed four grid steps on each beat of a tempo estimated independently for every file. This procedure made the median Nottingham tune \R{nott/lenratio} times longer than its score representation and mapped the same melody to different temporal grids across files. It also obscured near-duplicates: once the score grid was restored, \R{nott/moved} Nottingham tunes were found to share at least half of their 16-token patterns with a tune in another split.

A related receptive-field error also appeared in our first rerun. The initial exact-context predictor divided its nominal context approximately evenly across the two attention bands, which restricted effective copying to roughly half of the advertised context. The 336-token model therefore improved strongly on repeats up to 168 tokens back but much less on more distant repeats. Section~\ref{sec:causal} describes the corrected asymmetric banding scheme.

\textbf{A miscalibrated real-data baseline.} This issue affected the generation analysis of the earlier draft rather than the prediction curve. The draft reported a real-data distance floor of 0.36 based on a single set of 100 corpus sequences. That set lies at the 99.7th percentile of random 100-sequence draws and was not held out. Section~\ref{sec:gen} replaces it with a bootstrap reference band computed from held-out music. The broader evaluation lessons remain: generation metrics should be calibrated against the empirical distribution of real data~\cite{yang2020evaluation}; selecting candidates with an objective that shares terms with the evaluation metric can directly shift that metric; and tuning sampling temperature to minimize repetition can flatten a predictor to the point that trained and untrained weights produce similar outputs.

% ============================================================
\section{Discussion and Limitations}
\label{sec:discussion}
% ============================================================

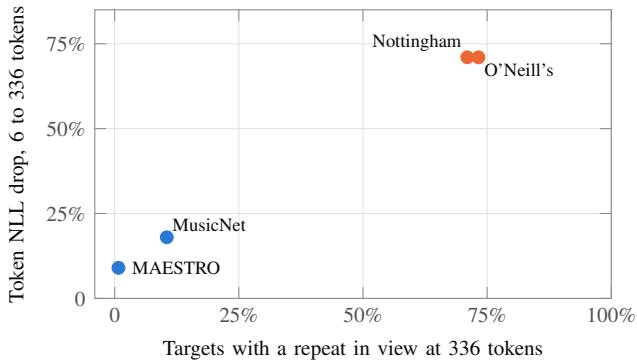
\begin{figure}[t]
\centering
% generated by scripts/make_paper_tables.py; do not edit
\definecolor{clat}{HTML}{2A78D6}
\definecolor{ctok}{HTML}{EB6834}
\definecolor{cink}{HTML}{0B0B0B}
\definecolor{cink2}{HTML}{52514E}
\definecolor{cmuted}{HTML}{898781}
\definecolor{cgrid}{HTML}{E1E0D9}
\begin{tikzpicture}
\begin{axis}[width=0.95\columnwidth, height=5.4cm, xmin=-4, xmax=100, ymin=0, ymax=85, clip=false,
  xtick={0,25,50,75,100}, xticklabels={0,25\%,50\%,75\%,100\%}, ytick={0,25,50,75}, yticklabels={0,25\%,50\%,75\%},
  xlabel={Targets with a repeat in view at 336 tokens}, ylabel={Token NLL drop, 6 to 336 tokens},
  grid=major, grid style={cgrid, line width=0.3pt}, axis line style={cmuted}, tick style={cmuted}, tick label style={font=\footnotesize, text=cink2}, label style={font=\footnotesize, text=cink}]
\node[anchor=south east, font=\scriptsize, text=cink, inner sep=2pt] at (axis cs:71.00,71.0) {Nottingham};
\addplot[only marks, mark=*, mark size=2.4pt, ctok] coordinates {(71.00,71.0)};
\node[anchor=north west, font=\scriptsize, text=cink, inner sep=2pt] at (axis cs:73.26,71.0) {O'Neill's};
\addplot[only marks, mark=*, mark size=2.4pt, ctok] coordinates {(73.26,71.0)};
\node[anchor=south west, font=\scriptsize, text=cink, inner sep=2pt] at (axis cs:10.49,18.0) {MusicNet};
\addplot[only marks, mark=*, mark size=2.4pt, clat] coordinates {(10.49,18.0)};
\node[anchor=west, xshift=3pt, font=\scriptsize, text=cink, inner sep=2pt] at (axis cs:0.76,9.0) {MAESTRO};
\addplot[only marks, mark=*, mark size=2.4pt, clat] coordinates {(0.76,9.0)};
\end{axis}
\end{tikzpicture}
\caption{The four corpora: share of test targets with a 16-token repeat in view at 336 tokens against the token predictor's NLL drop from 6 to 336 tokens. One point per corpus, so this is a description, not a test, and no line is fitted.}
\label{fig:scatter}
\end{figure}

\textbf{What determines the gain from long context.} On Nottingham and on the O'Neill's tunes long enough for evaluation, long-context gains track the availability of exact token-sequence repetitions; on MAESTRO they do not, and on MusicNet they do so only partially. Exact repetition is jointly determined by musical structure and representation. Our tokenization retains only the uppermost pitch on a fixed temporal grid and does not encode onsets, while both folk corpora are rendered scores in which written repeat signs are expanded note for note. A repeated section can therefore become exactly identical at the token level. By contrast, an expressive performance sampled every 125\,ms rarely reproduces a passage token for token. The difference between the folk corpora and MAESTRO should therefore be interpreted as a difference between rendered, quantized scores and expressive performance representations as much as a difference between musical genres. The copy baseline, which adds only three fitted interpolation weights to the 6-token predictor, recovers \R{nott/copybase/backoff/gain}\% of the long-context gain on Nottingham and \R{oneill/copybase/backoff/gain}\% on O'Neill's, indicating that most of the benefit in both corpora can be explained by retrieval of an exact earlier passage. Across the four evaluation sets, the proportion of targets with a repeat in view at 336 tokens (\R{nott/inview/336}\%, \R{oneill/inview/336}\%, \R{music/inview/336}\%, and \R{maes/inview/336}\% for Nottingham, O'Neill's, MusicNet, and MAESTRO) covaries descriptively with the corresponding token-predictor reductions (\R{nott/tok/drop}\%, \R{oneill/tok/drop}\%, \R{music/tok/drop}\%, and \R{maes/tok/drop}\%; Fig.~\ref{fig:scatter}). With only four corpora, this is descriptive evidence rather than a statistical test. Because the sweeps sample target positions differently across corpora (Section~\ref{sec:protocol}), we additionally compute repeat prevalence over every test position with at least 336 preceding tokens using a common rule; the resulting values, \R{common/nott}\%, \R{common/oneill}\%, \R{common/music}\%, and \R{common/maes}\%, closely match the sweep-target estimates. On MAESTRO, \R{maes/redshare/t/none}\% of the 6-to-336-token reduction comes from targets without an exact in-view repeat, and most of this improvement is already attained by 96 steps, or 12 seconds. We do not identify the information responsible for that residual gain.

\textbf{MAESTRO beyond 336 tokens.} Yi et al.~\cite{yi2026depth} report dependence on context beyond 1024 tokens for full-horizon MAESTRO models. Our results are compatible with that finding because we do not evaluate context beyond 336 tokens and use a substantially different setup, including a monophonic top-line representation, a 125\,ms grid, and approximately 2.5M parameters. Even within our tested range, the token predictor gains an additional \R{maes/boot/t/96-336} bits per token from 96 to 336 tokens, almost entirely on targets without an exact repeat.

\textbf{Latents.} The latent predictor also benefits from long context on Nottingham and is more stable than the token predictor across predictor seeds, although variation across autoencoder seeds is larger (Section~\ref{sec:aeseeds}). It benefits even when the earlier occurrence is encoded by a different latent code, indicating that the effect is not simply literal repetition of a previously observed code. Because latent NLL and token NLL are defined over different variables, however, their absolute values are incomparable and cannot establish which model better represents the music. Such a comparison would require a token-level likelihood for the latent model, which is not available under the factorized binary-code predictor used here.

\textbf{Scope.} The study is limited to four corpora, a monophonic top-line representation, context lengths up to 336 tokens, and predictors of approximately 2.5M parameters trained on at most a few million tokens. We therefore do not claim that 336 tokens is sufficient or optimal in general. Results in the range from 18 to 96 tokens are also sensitive to optimization: at 96 tokens, only about half of the Nottingham runs discover the copying solution within 200 epochs, so the observed curve may change with model capacity, data scale, or training duration. Our repeat analysis detects exact recurrence of a 16-token history; transposed repeats are treated separately (Section~\ref{sec:lag}), and approximate repetition, such as ornamented phrases or altered endings, is not measured. Much of the exact repetition in Nottingham and O'Neill's arises from written repeat signs expanded by the Nottingham MIDI renderer or by our O'Neill's preprocessing. Consequently, the second folk corpus tests whether the effect generalizes beyond Nottingham within rendered scores, not whether it holds for performed music. Recordings of the same tunes would be expected to exhibit fewer exact token-level repetitions, and collapsing written repeats would provide a direct test of how much the result depends on repeat expansion. MusicNet includes token predictors only and \R{music/pieces/testwin} test pieces with at least one evaluation window. The fixed 384-token window also excludes short tunes: \R{nott/sel/kept/n} of \R{nott/pieces/test} Nottingham test tunes and \R{oneill/sel/kept/n} of \R{oneill/pieces/test} O'Neill's test tunes contribute. Among all positions in the retained Nottingham tunes, \R{nott/sel/kept/repeat}\% have a 16-token history recurring within the previous 336 tokens, compared with \R{nott/sel/dropped/repeat}\% in the \R{nott/sel/dropped/n} excluded tunes; the corresponding O'Neill's values are \R{oneill/sel/kept/repeat}\% and \R{oneill/sel/dropped/repeat}\% across \R{oneill/sel/dropped/n} excluded tunes. The excluded tunes are shorter (median \R{nott/sel/dropped/median} and \R{oneill/sel/dropped/median} tokens versus \R{nott/sel/kept/median} and \R{oneill/sel/kept/median}), which mechanically reduces opportunities for recurrence. Restricting the comparison to the first 256 tokens largely removes the difference on Nottingham (\R{nott/sel/kept/repeat256}\% versus \R{nott/sel/dropped/repeat256}\%) but not on O'Neill's (\R{oneill/sel/kept/repeat256}\% versus \R{oneill/sel/dropped/repeat256}\%). Thus the O'Neill's evaluation subset is not only longer but also intrinsically more repetitive under this measure. The reported folk-corpus reductions therefore characterize tunes long enough to fill the evaluation window, with stronger selection toward repetition for O'Neill's.

% ============================================================
\section{Conclusion}
\label{sec:conclusion}
% ============================================================

Using source-disjoint splits with explicit near-duplicate control, non-overlapping latent receptive fields, and predictors whose receptive fields are matched to the intended context length, we find that the long-context gain on the two folk corpora is largely attributable to exact repetition in these rendered score representations. On Nottingham, increasing context from 6 to 336 tokens reduces token-predictor test NLL by \R{nott/tok/drop}\% (\R{nott/boot/t/6-336} bits per token) and latent-predictor NLL by \R{nott/lat/drop}\%; on the O'Neill's test tunes long enough for evaluation, the corresponding reductions are \R{oneill/tok/drop}\% (\R{oneill/boot/t/6-336} bits) and \R{oneill/lat/drop}\%. Most of these gains occur on targets whose history recurs earlier in the context. Targeted erasure of the earlier occurrence removes the gain, an explicit copy baseline recovers most of it, and a trained 336-token model loses the advantage when distant history is removed at test time. On MAESTRO and MusicNet, where 16-token repeats are much less frequent, the total gains within 336 tokens are smaller (\R{maes/boot/t/6-336} and \R{music/boot/t/6-336} bits), arise primarily on targets without such a repeat (\R{maes/redshare/t/none}\% and \R{music/redshare/t/none}\%), and are mostly attained by 96 tokens. On MAESTRO, even shorter exact-match retrieval explains little of the gain, with the copy baseline recovering only \R{maes/copybase/backoff/gain}\%. On MusicNet, the baseline recovers \R{music/copybase/backoff/gain}\%, indicating that exact matching could account for a larger fraction, although the Transformer itself obtains only \R{music/copybase/m/t336}\% of its total reduction on targets matched by the baseline. These conclusions are restricted to contexts of at most 336 tokens, the tested representations, and the stated training protocol; they do not determine a universal context requirement for music Transformers. The earlier 16-token saturation result coincided with four methodological failures: split leakage, target-overlapping latent receptive fields, an averaging predictor, and a per-file tempo grid. Context-length studies should explicitly test for each of these failure modes.

% ============================================================
\FloatBarrier

\appendix[Reproducibility Details]
\label{sec:repro}

\textbf{Availability.} The code, split manifests identifying every piece in each split, and outputs from every experimental run are available from the author on request. All reported numbers, tables, and figures are generated from these outputs by a single analysis script.

\textbf{Splits.} MAESTRO uses the official split. For the remaining corpora, pieces are grouped before splitting: one initial group per tune for Nottingham and O'Neill's, and one per composer and movement for MusicNet. Two pieces are linked when the number of distinct non-constant 16-token patterns they share is at least half the number of such patterns in the smaller piece; patterns occurring in more than 50 pieces are ignored. Linked groups are merged by connected components. Groups are sorted by name and permuted using NumPy's default random generator with seed 42; the first 70\% are assigned to training, the next 15\% to validation, and the remainder to test. For Nottingham and MusicNet, the near-duplicate rule is applied again after final tokenization; if a newly merged group spans multiple splits, all of its pieces are reassigned to the split containing the largest number of group members.

\begin{table}[t]
\centering
\caption{Attention bands and parameter counts of the predictors at each context length ($b_1 + b_2 - 1$ equals the context, in tokens for the token predictor and in latent steps for the latent predictor). The MAESTRO token predictors, with 89 tokens, have \R{params/tok89} parameters at 336 tokens.}
\label{tab:bands}
\resizebox{\columnwidth}{!}{% generated by scripts/make_paper_tables.py; do not edit
\begin{tabular}{rccc@{\hspace{1.2em}}rccc}
\toprule
\multicolumn{4}{c}{\textbf{Token predictor}} & \multicolumn{4}{c}{\textbf{Latent predictor}} \\
\cmidrule(lr){1-4}\cmidrule(lr){5-8}
$T$ & $b_1$ & $b_2$ & Parameters & $k$ & $b_1$ & $b_2$ & Parameters \\
\midrule
6 & 3 & 4 & 2{,}471{,}952 & 1 & 1 & 1 & 2{,}481{,}088 \\
18 & 9 & 10 & 2{,}473{,}104 & 3 & 2 & 2 & 2{,}481{,}280 \\
48 & 18 & 31 & 2{,}475{,}984 & 8 & 3 & 6 & 2{,}481{,}760 \\
96 & 18 & 79 & 2{,}480{,}592 & 16 & 3 & 14 & 2{,}482{,}528 \\
192 & 18 & 175 & 2{,}489{,}808 & 32 & 3 & 30 & 2{,}484{,}064 \\
336 & 18 & 319 & 2{,}503{,}632 & 56 & 3 & 54 & 2{,}486{,}368 \\
\bottomrule
\end{tabular}
}
\end{table}

\textbf{Predictors.} Table~\ref{tab:bands} lists the attention-band widths and parameter counts for every context length. Training seeds are 42, 7, and 0 for all main configurations; 0 to 9 for the fixed-budget runs in Section~\ref{sec:stability}; and 42, 1, and 2 for the three Nottingham autoencoders. Each seed initializes both the NumPy and PyTorch random-number generators.

\begin{table}[t]
\centering
\caption{Mixing weights of the copy baseline, fitted on validation windows.}
\label{tab:copyw}
% generated by scripts/make_paper_tables.py; do not edit
\begin{tabular}{lccc}
\toprule
 & \multicolumn{3}{c}{\textbf{History matched (tokens)}} \\
\cmidrule(lr){2-4}
\textbf{Corpus} & 32 & 16 & 8 \\
\midrule
Nottingham & 0.965 & 0.869 & 0.625 \\
O'Neill's & 0.969 & 0.890 & 0.733 \\
MusicNet & 0.902 & 0.866 & 0.760 \\
MAESTRO & 0.553 & 0.586 & 0.510 \\
\bottomrule
\end{tabular}

\end{table}

\textbf{Copy baseline.} For each target, the baseline searches the 336-token context for the nearest earlier match of the target's 32-token history, backing off to 16 tokens and then 8 tokens if necessary, and uses the first available match length. For a match of length $h$, it assigns token $v$ probability $\lambda_h \mathbf{1}[v=c] + (1-\lambda_h)p_6(v)$, where $c$ is the token following the matched history and $p_6$ is the 6-token predictor distribution. Each interpolation coefficient $\lambda_h$ is selected independently from a 1000-point grid spanning 0 to 0.999 to minimize NLL on validation targets matched at that history length (Table~\ref{tab:copyw}).

\textbf{Random edits.} Donor windows, control positions, and replacement pitches in the erase intervention are sampled with NumPy's default generator. The seed equals the context length for token predictors and 1000 plus the context length for latent predictors; additional independent draws add the draw index to this seed. The history-replacement intervention of Section~\ref{sec:ablate} uses one donor window per test window with seed 0.

\textbf{Generation.} Temperature search initially evaluates 0.3 to 0.8 in increments of 0.1 together with 1.0 for latent models, and 0.6 to 1.2 in increments of 0.1 for token models. The search is extended in 0.1 increments until the best validation temperature has an evaluated neighboring value on both sides.

\textbf{Software and hardware.} Experiments use Python 3.10.19, PyTorch 2.10.0, and NumPy 2.2.6 on macOS 26.6, using a single Apple M5 GPU with 24\,GB of memory through PyTorch's MPS backend. Execution on this backend is not bit-reproducible (Section~\ref{sec:protocol}).

\end{document}